\documentclass[natbib]{svjour3}

\usepackage{amsmath}
\usepackage{graphicx}
\usepackage{wasysym}
\usepackage{array}
\usepackage{lscape}

\allowdisplaybreaks

\newcolumntype{R}{>{$}r<{$}}
\newcolumntype{L}{>{$}l<{$}}
\newcolumntype{M}{R@{$\;$}L}

\begin{document}

\title{Linearized averaged resonant equations and their solution for dust
particles}

\author{Pavol P\'{a}stor}

\institute{Pavol P\'{a}stor \at
      Tekov Observatory, \\
      Sokolovsk\'{a} 21, 934~01, Levice, Slovak Republic \\
      \email{pavol.pastor@hvezdarenlevice.sk}
}

\date{}

\maketitle

\begin{abstract}
The averaged resonant equations of motion for the planar circular
restricted three-body problem are solved on the linearization basis
taking into account also non-gravitational effects. The averaged
resonant equations are derived from Lagrange's planetary equations
with additional Gauss's terms caused by the non-gravitational effects.
The time depending solution has the standard form with exponential,
quadratic, linear and constant terms. The existence of a rotational
symmetry in the action of the non-gravitational effects around the star
determines the order of a characteristic equation of the linearized
system. In the symmetrical case (order 3) the considered
non-gravitational effects are the stellar electromagnetic radiation and
the radial stellar wind (stellar radiation). In the asymmetrical case
(order 4) the stellar radiation and interstellar gas flow are
considered. It is investigated how well the linearization solution
describes real solution obtained from an equation of motion by a comparison
of the resonant libration frequency found analytically and numerically.
It is found that from initial values of the evolving orbital
parameters (semimajor axis, eccentricity, longitude of pericenter, and
resonant angular variable) in the averaged phase space the linearization
frequency depends most sensitively on the initial value of the resonant
angular variable. For small libration amplitudes of the resonant angular
variable the best match of the real libration frequency and the linearization
frequency is located approximately at the solution of the resonant
condition ($da / dt$ $=$ 0). If the initial averaged conditions are chosen
close to the solution of resonant condition, then the linearization
frequency for practically all simple oscillatory evolutions matches the real
libration frequency and the linearization solution very well approximates
the real evolution. The linearization results obtained for stationary
solutions are tested. In the planar circular restricted Sun-Neptune-dust
problem with the solar radiation and the interstellar gas flow the solutions
of resonant condition are practically independent on the longitude
of perihelion.

\end{abstract}

\section{Introduction}
\label{sec:intro}

The gravity of a star and a planet that move according to a solution
of the two body problem disturbs the motion of a body with negligible
mass (restricted three-body problem). The dynamics of the body with
negligible mass includes in this case also the so called mean motion
resonances. In a mean motion resonance a ratio of orbital periods of the two
minor bodies oscillates near a ratio of two natural numbers. The motion
of the body with negligible mass in the mean motion resonance cannot be
solved completely even in the planar case (when the motions are confined to
one plane). Several approximative solutions for an averaged problem can be
found in the literature. The behavior predicted by the averaged solutions
depends on the period of averaging. After the averaging over a synodic period
oscillations in the evolution of semimajor axis should be present.
The oscillations are also present if the averaged solution is determined
using Fourier series expansion of the disturbing function with considered
single resonant term \citep[e.g.][]{greenberg,mude}. After the averaging
over the libration period the semimajor axis should be constant.

\citet{greenberg} substituted the truncated averaged Fourier series
expansion of the disturbing function in the time derivatives of orbital
elements given by Lagrange's planetary equations. Lagrange's planetary
equations for the planar circular restricted three-body problem (PCRTBP)
including a tidal dissipation\footnote{The tidal dissipation was
introduced as a migration of the perturbing body in a circular orbit.} were
solved simultaneously using several approximations. Another example
of the time dependence obtained using the truncated Fourier series
of the disturbing function can be found in \citet{mude}. By individual
time integration of Lagrange's planetary equations in the PCRTBP with
substituted single resonant term from the Fourier series they obtained
dependencies of orbital elements on time. They assumed that the only
time-varying quantities in the equations are in the trigonometric arguments
of the resonant term and that the longitude of perihelion increases
linearly with time at a constant rate determined by secular theory. Evolutions
obtained from Fourier series expansion of the disturbing function will be
no more discussed in this paper.

The evolutions averaged over the libration period are commonly used
for the description of the long term evolution of dust particles captured
in the mean motion resonances with the planet. The dust particles
captured in a neighborhood of the Earth's orbit were predicted by
\citet{shepherd} and observed in the infrared light by the satellites
{\it IRAS} \citep{IRAS} and {\it COBE} \citep{COBE}. The dust particles are
significantly influenced by non-gravitational effects. When the resonant
dust particles are under the action of the Poynting--Roberson (PR) effect
\citep{poynting,robertson,burns,klacka2004,icarus} and a radial
solar wind \citep{covsw}, then the evolution of eccetrtricity
averaged over the libration period shows a sorted monotonic
behavior. Properties of this behavior were investigated in some depth
by \citet{WJ,FM2,gomes95,LZJ}; and others. After the averaging over
the libration period these particles follow the eccentricity
evolution described by a first order differential equation derived
in \citet{LZ1997}. In \citet{LZ1997} authors expanded the derived equation
for the evolution of eccentricity to the second order in the eccentricity.
The obtained equation was solved for a time dependence valid for small
eccentricities. The time dependence of the eccentricity is frequently
used \citep[see e.g.][]{MM,DM,krivov}. Results obtained after
the averaging over the libration period will be not taken a step further
in this paper.

In \citet{FM2} the equations of motion of a dust particle
captured in a mean motion resonance in the PCRTBP with the PR effect were
written in a near canonical form. \citet{FM2} transformed the near canonical
equations to a system of equations suitable for the search of stationary
points and averaged them over a synodic period (averaged resonant
equations). \citet{FM2} linearized the averaged resonant equations
around chosen stationary point and solved obtained characteristic
equation of the system in order to determine a stability of the stationary
points. This method was used for stability tests of the stationary
points in the PCRTBP with the PR effect also by \citet{sidnes}. In \citet{kh}
periodic motions in a reference frame rotating with the planet were found
to exist at each of such stationary points obtained from the averaged resonant
equations. \citet{triangular} found stationary points in the circular-planar,
spatial-circular, elliptic-planar and spatial-elliptic restricted three-body
problem with the PR effect for the dust particles captured in the mean motion
1/1 resonance with the planet \citep[see also][]{points}. The stability
of found stationary points was investigated using the linearization
of the equations of motion written in Delaunay variables and averaged over
the orbital period.

In this paper we derive the solution of linearized averaged resonant
equation for the PCRTBP with non-gravitational effect in general form.
The derived solution should be valid for any mean motion resonance.
The non-gravitational effects with and without the rotational symmetry around
the star will be considered separately. The linearization is usually
used in the literature to investigate a stability, and to search for
linearization frequencies, but how well the time depending solution describes
real resonant librations was not yet presented in the literature.
We find that the linearization frequencies significantly depend on
the initial conditions in the averaged phase space even for one libration.
We show how the solution should be applied for the sake of best
description of almost all evolutions with simple oscillations in
the mean motion resonances. Frequencies for periodic solutions in exterior
mean motion 6/5, 7/6, 8/7, and 9/8 resonances with the Earth in a circular
orbit will be determined. The applicability will be investigated when
the non-gravitational effects are the PR effect, radial solar wind and
interstellar gas flow.

\section{Averaged resonant equations}
\label{sec:resonantequations}

For the study of a specific mean motion resonance it is convenient to define
a resonant angular variable \citep[e.g.][]{greenberg,FM1,FM2,gomes95}
\begin{equation}\label{sigma}
\sigma = \frac{p + q}{q} \lambda_{\text{P}} - s \lambda - \tilde{\omega} ~,
\end{equation}
here $p$ and $q$ are two integers (resonant numbers), $\lambda_{\text{P}}$
is the mean longitude of the planet in a circular orbit, $\lambda$ is the mean
longitude of the dust particle, $\tilde{\omega}$ is the longitude
of pericenter, and $s$ $=$ $p / q$. In what follows we will need also
the time derivative of the resonant angular variable. The mean longitude
of the planet increases linearly with the time $t$ from its initial
value $\lambda_{\text{P}0}$ with a constant slope equal to the mean motion
of the planet $n_{\text{P}}$ ($\lambda_{\text{P}}$ $=$ $n_{\text{P}} t$ $+$
$\lambda_{\text{P}0}$). We define an angle $\sigma_{\text{b}}$ so that
the mean anomaly of the dust particle can be computed from the relation
$M$ $=$ $n t$ $+$ $\sigma_{\text{b}}$ using the mean motion of the particle
$n$ and the time \citep{fund}. The mean motion of the particle
with the negligible mass is given by the third Kepler's law $n$ $=$
$\sqrt{\mu / a^{3}}$, here $\mu$ $=$ $G_{0} M_{\star}$, $G_{0}$ is
the gravitational constant, $M_{\star}$ is the mass of the star, and $a$
is the semimajor axis of the particle's orbit. The mean longitude of the dust
particle is according to the definitions above $\lambda$ $=$ $M$ $+$
$\tilde{\omega}$ $=$ $n t$ $+$ $\sigma_{\text{b}}$ $+$ $\tilde{\omega}$.
In the mean motion resonance $\sigma$ is librating rather than
circulating. For the time derivative of $\sigma$ we have
\begin{equation}\label{dsigmadt}
\frac{d \sigma}{dt} = \frac{p + q}{q} n_{\text{P}} - s n -
s \left ( \frac{d \sigma_{\text{b}}}{dt} + t \frac{dn}{dt} +
\frac{d \tilde{\omega}}{dt} \right ) -
\frac{d \tilde{\omega}}{dt} ~.
\end{equation}

Short periodic variations in the evolution during the mean motion resonance
can be ignored in the most practical cases. This can be done effectively
by an averaging over a synodic period. The synodic period is determined
by a difference in mean longitudes of the planet and the particle, and
the order of resonance $q$ in the angle variable
\begin{equation}\label{cyclic}
\sigma_{\text{T}} = \frac{\lambda - \lambda_{\text{P}}}{q} ~.
\end{equation}
The difference between $\sigma_{\text{T}}$ at time zero and $\sigma_{\text{T}}$
after one synodic period is equal to $2 \pi$.

The orbital evolution of a dust particle is significantly influenced
also by non-gravitational effects. For the non-gravitational effects
that slowly vary the dust particle's orbit the long term (secular) orbital
evolution can be described by the averaged time derivatives of the orbital
elements. The averaged time derivatives can be calculated using Gauss's
perturbation equations of celestial mechanics \citep[e.g.][]{danby,mude}.
The secular time derivatives of the orbital elements caused by the planet
can be calculated using Lagrange's planetary equations averaged over
the synodic period \citep{brocle,danby}. After the averaging we can sum
Lagrange's planetary equations and Gauss's perturbation equations in
order to obtain the system of equations describing the secular orbital
evolution of the dust particle. In the planar case the equations are
\begin{align}\label{lagrange}
\frac{da}{dt} &= \frac{2 a}{L}
      \frac{\partial R}{\partial \sigma_{\text{b}}} +
      \left ( \frac{da}{dt} \right )_{\text{EF}} ~,
\notag \\
\frac{de}{dt} &= \frac{\alpha^{2}}{L e}
      \frac{\partial R}{\partial \sigma_{\text{b}}} -
      \frac{\alpha}{L e}
      \frac{\partial R}{\partial \tilde{\omega}} +
      \left ( \frac{de}{dt} \right )_{\text{EF}} ~,
\notag \\
\frac{d \tilde{\omega}}{dt} &= \frac{\alpha}{L e}
      \frac{\partial R}{\partial e} +
      \left ( \frac{d \tilde{\omega}}{dt} \right )_{\text{EF}} ~,
\notag \\
\frac{d \sigma_{\text{b}}}{dt} + t \frac{dn}{dt} &= - \frac{2 a}{L}
      \frac{\partial R}{\partial^{\star} a} -
      \frac{\alpha^{2}}{L e}
      \frac{\partial R}{\partial e} +
      \left ( \frac{d \sigma_{\text{b}}}{dt} +
      t \frac{dn}{dt} \right )_{\text{EF}} ~.
\end{align}
Here $L$ $=$ $\sqrt{\mu a}$, $e$ is the eccentricity, and $\alpha$ $=$
$\sqrt{1 - e^{2}}$. $R$ is the disturbing function of the PCRTBP
\begin{equation}\label{DF}
R = G_{0} M_{\text{P}} \left ( \frac{1}{\vert \vec{r} -
\vec{r}_{\text{P}} \vert} -
\frac{\vec{r} \cdot \vec{r}_{\text{P}}}{r_{\text{P}}^{3}} \right )
\end{equation}
with its partial derivatives in Eqs. (\ref{lagrange}) averaged over
the synodic period. In the disturbing function: $M_{\text{P}}$ is the mass
of the planet, $\vec{r}_{\text{P}}$ is the position vector of the planet
with respect to the star, $r_{\text{P}}$ $=$ $\vert \vec{r}_{\text{P}} \vert$,
and $\vec{r}$ is the position vector of the dust particle with respect
to the star. The subscript EF in Eqs. (\ref{lagrange}) denotes
the terms that are caused by the non-gravitational effects only.
$\partial R / \partial^{\star} a$ in the last equation in
Eqs. (\ref{lagrange}) denotes that the partial derivative of the disturbing
function with respect to the semimajor axis is calculated with an assumption
that the mean motion of the particle $n$ is not a function of the semimajor
axis. This can be shown in the un-averaged phase space since we have
\begin{align}\label{pstar}
\left ( \frac{d \sigma_{\text{b}}}{dt} \right )_{\text{G}} &= - \frac{2 a}{L}
      \frac{\partial R}{\partial a} -
      \frac{\alpha^{2}}{L e}
      \frac{\partial R}{\partial e} = - \frac{2 a}{L}
      \frac{\partial R}{\partial^{\star} a} -
      \frac{\alpha^{2}}{L e}
      \frac{\partial R}{\partial e} -
      \frac{2 a}{L}
      \frac{\partial R}{\partial \sigma_{\text{b}}}
      t \frac{\partial n}{\partial a}
\notag \\
& = - \frac{2 a}{L}
      \frac{\partial R}{\partial^{\star} a} -
      \frac{\alpha^{2}}{L e}
      \frac{\partial R}{\partial e} -
      \left ( t \frac{dn}{dt} \right )_{\text{G}} ~.
\end{align}
The subscript G denotes the terms that are caused by the gravitation
only. In Eqs. (\ref{pstar}) we have substituted also the un-averaged
gravitational part of the first equation in Eqs. (\ref{lagrange}).
Using Eqs. (\ref{pstar}) averaged over the synodic period, we can obtain
the last equation in Eqs. (\ref{lagrange}) as follows
\begin{align}\label{terms}
\frac{d \sigma_{\text{b}}}{dt} + t \frac{dn}{dt} &=
      \left ( \frac{d \sigma_{\text{b}}}{dt} \right )_{\text{G}} +
      \left ( \frac{d \sigma_{\text{b}}}{dt} \right )_{\text{EF}} +
      \left ( t \frac{dn}{dt} \right )_{\text{G}} +
      \left ( t \frac{dn}{dt} \right )_{\text{EF}}
\notag \\
& = - \frac{2 a}{L}
      \frac{\partial R}{\partial^{\star} a} -
      \frac{\alpha^{2}}{L e}
      \frac{\partial R}{\partial e} +
      \left ( \frac{d \sigma_{\text{b}}}{dt} +
      t \frac{dn}{dt} \right )_{\text{EF}} ~.
\end{align}
Equations (\ref{lagrange}) already describe the secular evolution of the dust
particle captured in the mean motion resonance and simultaneously affected
by the non-gravitational effects.

The secular evolution in a specific resonance given by the resonant numbers
$p$ and $q$ cannot be easily seen in Eqs. (\ref{lagrange}). In order to study
the resonances we transform Eqs. (\ref{lagrange}) as follows.
Equation (\ref{dsigmadt}) can be averaged over the synodic period. If we
use in the averaged result the last two equations in Eqs. (\ref{lagrange}),
then we get
\begin{align}\label{fourth}
\frac{d \sigma}{dt} = {} & - \frac{\alpha}{L e}
      \left [ 1 + s \left ( 1 - \alpha \right ) \right ]
      \frac{\partial R}{\partial e} +
      \frac{2 s a}{L}
      \frac{\partial R}{\partial^{\star} a} +
      n_{\text{P}} \frac{p + q}{q} - n s
\notag \\
& - \frac{p + q}{q}
      \left ( \frac{d \tilde{\omega}}{dt} \right )_{\text{EF}} -
      s \left ( \frac{d \sigma_{\text{b}}}{dt} +
      t \frac{dn}{dt} \right )_{\text{EF}} ~.
\end{align}
The partial derivatives of the disturbing function $R$ averaged over
the synodic period are not functions of $\tilde{\omega}$ they are only
functions of $a$, $e$, and $\sigma$. Between averaged
partial derivatives of the disturbing function $R$ the following relations
hold
\begin{align}\label{relations}
\frac{\partial R}{\partial \sigma_{\text{b}}} &= - s
\frac{\partial R}{\partial \sigma} ~,
\notag \\
\frac{\partial R}{\partial \tilde{\omega}} &= - \frac{p + q}{q}
\frac{\partial R}{\partial \sigma} ~.
\end{align}
We can use Eqs. (\ref{relations}) in the first two equations in
the system of equations given by Eqs. (\ref{lagrange}). The last
equation in Eqs. (\ref{lagrange}) can be replaced with equivalent
Eq. (\ref{fourth}). By this we obtain a system that enables
the study of the secular evolution of the dust particle captured in
the specific mean motion resonance given by the resonant numbers
$p$ and $q$ under the action of the non-gravitational effects
\begin{align}\label{evolution}
\frac{da}{dt} = {} & - \frac{2 s a}{L}
      \frac{\partial R}{\partial \sigma} +
      \left ( \frac{da}{dt} \right )_{\text{EF}} ~,
\notag \\
\frac{de}{dt} = {} & \frac{\alpha}{L e}
      \left [ 1 + s \left ( 1 - \alpha \right ) \right ]
      \frac{\partial R}{\partial \sigma} +
      \left ( \frac{de}{dt} \right )_{\text{EF}} ~,
\notag \\
\frac{d \tilde{\omega}}{dt} = {} & \frac{\alpha}{L e}
      \frac{\partial R}{\partial e} +
      \left ( \frac{d \tilde{\omega}}{dt} \right )_{\text{EF}} ~,
\notag \\
\frac{d \sigma}{dt} = {} & - \frac{\alpha}{L e}
      \left [ 1 + s \left ( 1 - \alpha \right ) \right ]
      \frac{\partial R}{\partial e} +
      \frac{2 s a}{L} \frac{\partial R}{\partial^{\star} a} +
      n_{\text{P}} \frac{p + q}{q} - n s
\notag \\
& - \frac{p + q}{q}
      \left ( \frac{d \tilde{\omega}}{dt} \right )_{\text{EF}} -
      s \left ( \frac{d \sigma_{\text{b}}}{dt} +
      t \frac{dn}{dt} \right )_{\text{EF}} ~.
\end{align}
Equations (\ref{evolution}) still valid also close to the zero
eccentricity. Singularities in the eccentricity reflect noncontinuous
behavior in the evolutions at the zero eccentricity. For example, it is
possible that a decrease of the eccentricity does change suddenly to an
increase at the zero eccentricity. Singularities in eccentricities reflect
also definitions of the orbital elements. For example, the longitude
of pericenter is not defined at the zero eccentricity. The last term
in Eqs. (\ref{evolution}) despite of its complicated appearance can be
straightforwardly obtained using \citep{fund}
\begin{align}\label{passage}
\left ( \frac{d \sigma_{\text{b}}}{dt} +
      t \frac{dn}{dt} \right )_{\text{EF}} =
      \left ( \frac{dM}{dt} - n \right )_{\text{EF}} = {} &
      \biggl \langle \frac{1 - e^{2}}{n a} \biggl [ a_{\text{R}} \left (
      \frac{\cos f}{e} - \frac{2}{1 + e \cos f} \right )
\notag \\
& - a_{\text{T}} \frac{\sin f}{e} \frac{2 + e \cos{f}}{1 + e \cos f} \biggr ]
      \biggr \rangle ~,
\end{align}
where $f$ is the true anomaly, $a_{\text{R}}$ and $a_{\text{T}}$ are
the radial and transversal components of the acceleration caused by
the non-gravitational effects. The angle brackets in Eq. (\ref{passage})
denote an averaging over one orbital period $T$ \citep[see e.g.][]{klacka2004}
\begin{equation}\label{average}
\langle g \rangle = \frac{1}{T} \int_{0}^{T} g (t) ~dt =
\frac{1}{T} \int_{0}^{2 \pi} g (f) ~\frac{dt}{df} ~df =
\frac{1}{2 \pi a^{2} \sqrt{1 - e^{2}}} \int_{0}^{2 \pi} r^{2} ~g (f) ~df .
\end{equation}

The system of equations given by Eqs. (\ref{evolution}) is different
from systems considered in \citet{FM2} and \citet{sidnes}.
Equations (8) in \citet{FM2} are equivalent with Eqs. (\ref{evolution}) if
\begin{align}\label{FM-case}
\left ( \frac{d \tilde{\omega}}{dt} \right )_{\text{EF}} = {} & 0 ~,
\notag \\
\left ( \frac{d \sigma_{\text{b}}}{dt} +
      t \frac{dn}{dt} \right )_{\text{EF}} = {} & 0 ~.
\end{align}
Similarly as for the system of equations in \citet{FM2} Eqs. (18) in
\citet{sidnes} include only the non-gravitational effects for which
Eqs. (\ref{FM-case}) hold. \citet{sidnes} have only three equations in
the system. The evolution of the longitude of pericenter is ignored in
\citet{sidnes}. The equations of motion in this paper
(Eqs. \ref{evolution}) are usable also for the non-gravitational effects
that can have non-zero secular variations on the left-hand side
of Eqs. (\ref{FM-case}). This property make them usable also for
the non-gravitational effects acting without a rotational symmetry around
the star \citep[see Appendix A in][]{kh}.

\section{Linearization of averaged resonant equations}
\label{sec:linearization}

No general method exists for solving nonlinear differential
equations in the system Eqs. (\ref{evolution}). If another method
(use of nonlinear coordinate transformations, Lie transformations, etc.)
does not allow go further, then the best that can be accomplished
\citep{ames} is to study a linearization based upon initial conditions
for the function and its derivatives. In the vicinity
of an initial point $a_{0}$, $e_{0}$, $\tilde{\omega}_{0}$,
and $\sigma_{0}$ we use notation
\begin{align}\label{deltas}
\delta_{a} &= a - a_{0} ~,
\notag \\
\delta_{e} &= e - e_{0} ~,
\notag \\
\delta_{\tilde{\omega}} &= \tilde{\omega} - \tilde{\omega}_{0} ~,
\notag \\
\delta_{\sigma} &= \sigma - \sigma_{0} ~.
\end{align}
The time will be measured from an initial time $t_{0}$ $=$ 0. Hence
\begin{equation}\label{time}
\delta_{t} = t ~.
\end{equation}
On the left-hand side of Eqs. (\ref{evolution}) we substitute identities
from Eqs. (\ref{deltas}). Then for example, the time derivative
of the semimajor axis can be written as follows
\begin{equation}\label{substitution}
\frac{da}{dt} = \frac{d}{dt} (\delta_{a} + a_{0}) = \frac{d \delta_{a}}{dt} ~.
\end{equation}
The linearization of averaged resonant equations in the used notation is
\begin{alignat}{11}\label{linear}
\frac{d \delta_{a}}{dt} {} & = {} & {} & {} &
      {} & \biggl [ \frac{\partial}{\partial^{\star} a} {} &
      {} & \left ( \frac{da}{dt} \right ) \biggr ]_{0} {} &
      {} & \delta_{a} {} & {} & + {} &
      {} & \biggl [ \frac{\partial}{\partial e} {} &
      {} & \left ( \frac{da}{dt} \right ) \biggr ]_{0} {} &
      {} & \delta_{e} {} & {} & + {} &&
      \biggl [ \frac{\partial}{\partial \tilde{\omega}}
      \left ( \frac{da}{dt} \right ) \biggr ]_{0}
      \delta_{\tilde{\omega}}
\notag \\
{} & {} & {} & + {} &
      {} & \biggl [ \frac{\partial}{\partial \sigma} {} &
      {} & \left ( \frac{da}{dt} \right ) \biggr ]_{0} {} &
      {} & \delta_{\sigma} {} & {} & + {} &
      {} & \biggl [ \frac{\partial}{\partial t} {} &
      {} & \left ( \frac{da}{dt} \right ) \biggr ]_{0} {} &
      {} & t {} & {} & + {} &&
      \left ( \frac{da}{dt} \right )_{0} ~,
\notag \\
\frac{d \delta_{e}}{dt} {} & = {} & {} & {} &
      {} & \biggl [ \frac{\partial}{\partial^{\star} a} {} &
      {} & \left ( \frac{de}{dt} \right ) \biggr ]_{0} {} &
      {} & \delta_{a} {} & {} & + {} &
      {} & \biggl [ \frac{\partial}{\partial e} {} &
      {} & \left ( \frac{de}{dt} \right ) \biggr ]_{0} {} &
      {} & \delta_{e} {} & {} & + {} &&
      \biggl [ \frac{\partial}{\partial \tilde{\omega}}
      \left ( \frac{de}{dt} \right ) \biggr ]_{0}
      \delta_{\tilde{\omega}}
\notag \\
{} & {} & {} & + {} &
      {} & \biggl [ \frac{\partial}{\partial \sigma} {} &
      {} & \left ( \frac{de}{dt} \right ) \biggr ]_{0} {} &
      {} & \delta_{\sigma} {} & {} & + {} &
      {} & \biggl [ \frac{\partial}{\partial t} {} &
      {} & \left ( \frac{de}{dt} \right ) \biggr ]_{0} {} &
      {} & t {} & {} & + {} &&
      \left ( \frac{de}{dt} \right )_{0} ~,
\notag \\
\frac{d \delta_{\tilde{\omega}}}{dt} {} & = {} & {} & {} &
      {} & \biggl [ \frac{\partial}{\partial^{\star} a} {} &
      {} & \left ( \frac{d \tilde{\omega}}{dt} \right )
      \biggr ]_{0} {} &
      {} & \delta_{a} {} & {} & + {} &
      {} & \biggl [ \frac{\partial}{\partial e} {} &
      {} & \left ( \frac{d \tilde{\omega}}{dt} \right )
      \biggr ]_{0} {} &
      {} & \delta_{e} {} & {} & + {} &&
      \biggl [ \frac{\partial}{\partial \tilde{\omega}}
      \left ( \frac{d \tilde{\omega}}{dt} \right ) \biggr ]_{0}
      \delta_{\tilde{\omega}}
\notag \\
{} & {} & {} & + {} &
      {} & \biggl [ \frac{\partial}{\partial \sigma} {} &
      {} & \left ( \frac{d \tilde{\omega}}{dt} \right )
      \biggr ]_{0} {} &
      {} & \delta_{\sigma} {} & {} & + {} &
      {} & \biggl [ \frac{\partial}{\partial t} {} &
      {} & \left ( \frac{d \tilde{\omega}}{dt} \right )
      \biggr ]_{0} {} &
      {} & t {} & {} & + {} &&
      \left ( \frac{d \tilde{\omega}}{dt} \right )_{0} ~,
\notag \\
\frac{d \delta_{\sigma}}{dt} {} & = {} & {} & {} &
      {} & \biggl [ \frac{\partial}{\partial^{\star} a} {} &
      {} & \left ( \frac{d \sigma}{dt} \right ) \biggr ]_{0} {} &
      {} & \delta_{a} {} & {} & + {} &
      {} & \biggl [ \frac{\partial}{\partial e} {} &
      {} & \left ( \frac{d \sigma}{dt} \right ) \biggr ]_{0} {} &
      {} & \delta_{e} {} & {} & + {} &&
      \biggl [ \frac{\partial}{\partial \tilde{\omega}}
      \left ( \frac{d \sigma}{dt} \right ) \biggr ]_{0}
      \delta_{\tilde{\omega}}
\notag \\
{} & {} & {} & + {} &
      {} & \biggl [ \frac{\partial}{\partial \sigma} {} &
      {} & \left ( \frac{d \sigma}{dt} \right ) \biggr ]_{0} {} &
      {} & \delta_{\sigma} {} & {} & + {} &
      {} & \biggl [ \frac{\partial}{\partial t} {} &
      {} & \left ( \frac{d \sigma}{dt} \right ) \biggr ]_{0} {} &
      {} & t {} & {} & + {} &&
      \left ( \frac{d \sigma}{dt} \right )_{0} ~,
\end{alignat}
$\partial^{\star} a$ is used here since after the averaging
dependencies on $M$ in the terms with the disturbing function are lost.
We calculate the derivatives with respect to the semimajor axis in
the derivatives of disturbing function during the averaging at a given mean
anomaly $M$ regardless of $M$ variation caused by the semimajor axis.
This holds also for the eccentricity since $\partial M / \partial e$ $=$ 0
(but $\partial f / \partial e$ $\neq$ 0). The partial derivatives with
respect to the time are usable only for time variations of the solved
problem that are negligible during the averaging over the synodic period.
The terms with the subscript 0 on the right-hand sides in Eqs. (\ref{linear})
are constant therefore we can simply write
\begin{alignat}{14}\label{oscillations}
{} & \dot{\delta}_{a} {} & {} & = {} &
      {} & A_{\text{c}} \delta_{a} {} & {} & + {} &
      {} & B_{\text{c}} \delta_{e} {} & {} & + {} &
      {} & C_{\text{c}} \delta_{\tilde{\omega}} {} & {} & + {} &
      {} & D_{\text{c}} \delta_{\sigma} {} & {} & + {} &
      {} & E_{\text{c}} t {} & {} & + {} & {} & F {} & ~, {} &
\notag \\
{} & \dot{\delta}_{e} {} & {} & = {} &
      {} & G_{\text{c}} \delta_{a} {} & {} & + {} &
      {} & H_{\text{c}} \delta_{e} {} & {} & + {} &
      {} & I_{\text{c}} \delta_{\tilde{\omega}} {} & {} & + {} &
      {} & J_{\text{c}} \delta_{\sigma} {} & {} & + {} &
      {} & K_{\text{c}} t {} & {} & + {} & {} & L {} & ~, {} &
\notag \\
{} & \dot{\delta}_{\tilde{\omega}} {} & {} & = {} &
      {} & M_{\text{c}} \delta_{a} {} & {} & + {} &
      {} & N_{\text{c}} \delta_{e} {} & {} & + {} &
      {} & O_{\text{c}} \delta_{\tilde{\omega}} {} & {} & + {} &
      {} & P_{\text{c}} \delta_{\sigma} {} & {} & + {} &
      {} & Q_{\text{c}} t {} & {} & + {} & {} & R {} & ~, {} &
\notag \\
{} & \dot{\delta}_{\sigma} {} & {} & = {} &
      {} & S_{\text{c}} \delta_{a} {} & {} & + {} &
      {} & T_{\text{c}} \delta_{e} {} & {} & + {} &
      {} & U_{\text{c}} \delta_{\tilde{\omega}} {} & {} & + {} &
      {} & V_{\text{c}} \delta_{\sigma} {} & {} & + {} &
      {} & W_{\text{c}} t {} & {} & + {} & {} & X {} & ~. {} &
\end{alignat}
This system describes solution of system Eqs. (\ref{evolution}) during
a short time interval after the initial time $t_{0}$ $=$ 0
(see Appendix \ref{app:coefficients}).

It is possible to obtain an equation for one chosen variation by
an elimination of the remaining variations using all equations
in the system. The obtained equations for the separated variations are
(see Appendix \ref{app:separation})
\begin{alignat}{15}\label{complete}
{} & \ddddot{\delta}_{a} {} &
      {} & + {} &
      {} & \Lambda_{a 3} ~\dddot{\delta}_{a} {} &
      {} & + {} &
      {} & \Lambda_{a 2} ~\ddot{\delta}_{a} {} &
      {} & + {} &
      {} & \Lambda_{a 1} ~\dot{\delta}_{a} {} &
      {} & + {} &
      {} & \Lambda_{a 0} ~\delta_{a} {} &
      {} & + {} &
      {} & \Lambda_{a t} ~t {} &
      {} & + {} &
      {} & \Lambda_{a} {} & {} & = {} & {} & 0 ~,
\notag \\
{} & \ddddot{\delta}_{e} {} &
      {} & + {} &
      {} & \Lambda_{e 3} ~\dddot{\delta}_{e} {} &
      {} & + {} &
      {} & \Lambda_{e 2} ~\ddot{\delta}_{e} {} &
      {} & + {} &
      {} & \Lambda_{e 1} ~\dot{\delta}_{e} {} &
      {} & + {} &
      {} & \Lambda_{e 0} ~\delta_{e} {} &
      {} & + {} &
      {} & \Lambda_{e t} ~t {} &
      {} & + {} &
      {} & \Lambda_{e} {} & {} & = {} & {} & 0 ~,
\notag \\
{} & \ddddot{\delta}_{\tilde{\omega}} {} &
      {} & + {} &
      {} & \Lambda_{\tilde{\omega} 3} ~\dddot{\delta}_{\tilde{\omega}} {} &
      {} & + {} &
      {} & \Lambda_{\tilde{\omega} 2} ~\ddot{\delta}_{\tilde{\omega}} {} &
      {} & + {} &
      {} & \Lambda_{\tilde{\omega} 1} ~\dot{\delta}_{\tilde{\omega}} {} &
      {} & + {} &
      {} & \Lambda_{\tilde{\omega} 0} ~\delta_{\tilde{\omega}} {} &
      {} & + {} &
      {} & \Lambda_{\tilde{\omega} t} ~t {} &
      {} & + {} &
      {} & \Lambda_{\tilde{\omega}} {} & {} & = {} & {} & 0 ~,
\notag \\
{} & \ddddot{\delta}_{\sigma} {} &
      {} & + {} &
      {} & \Lambda_{\sigma 3} ~\dddot{\delta}_{\sigma} {} &
      {} & + {} &
      {} & \Lambda_{\sigma 2} ~\ddot{\delta}_{\sigma} {} &
      {} & + {} &
      {} & \Lambda_{\sigma 1} ~\dot{\delta}_{\sigma} {} &
      {} & + {} &
      {} & \Lambda_{\sigma 0} ~\delta_{\sigma} {} &
      {} & + {} &
      {} & \Lambda_{\sigma t} ~t {} &
      {} & + {} &
      {} & \Lambda_{\sigma} {} & {} & = {} & {} & 0 ~.
\end{alignat}
In the used notation for the constants in Eqs. (\ref{complete}) we obtain
\begin{alignat}{9}\label{lambda3}
\Lambda_{3} = \Lambda_{a 3} = \Lambda_{e 3} = \Lambda_{\tilde{\omega} 3} =
      \Lambda_{\sigma 3} = {} & - {} &
      {} & A_{\text{c}} {} & {} & - {} & {} & H_{\text{c}} {} & {} & - {} &
      {} & O_{\text{c}} {} & {} & - {} & {} & V_{\text{c}} ~, {} & {} &
\end{alignat}
\begin{alignat}{7}\label{lambda2}
\Lambda_{2} = \Lambda_{a 2} = \Lambda_{e 2} = \Lambda_{\tilde{\omega} 2} =
      \Lambda_{\sigma 2} = {} & {} & {} & \left | \begin{array}{cc}
      A_{\text{c}} & B_{\text{c}} \\
      G_{\text{c}} & H_{\text{c}} \\
      \end{array} \right | {} & {} & + {} & {} & \left | \begin{array}{cc}
      A_{\text{c}} & C_{\text{c}} \\
      M_{\text{c}} & O_{\text{c}} \\
      \end{array} \right | {} & {} & + {} & {} & \left | \begin{array}{cc}
      A_{\text{c}} & D_{\text{c}} \\
      S_{\text{c}} & V_{\text{c}} \\
      \end{array} \right | {} & {} &
\notag \\
{} & + {} & {} & \left | \begin{array}{cc}
      H_{\text{c}} & I_{\text{c}} \\
      N_{\text{c}} & O_{\text{c}} \\
      \end{array} \right | {} & {} & + {} & {} & \left | \begin{array}{cc}
      H_{\text{c}} & J_{\text{c}} \\
      T_{\text{c}} & V_{\text{c}} \\
      \end{array} \right | {} & {} & + {} & {} & \left | \begin{array}{cc}
      O_{\text{c}} & P_{\text{c}} \\
      U_{\text{c}} & V_{\text{c}} \\
      \end{array} \right | ~, {} & {} &
\end{alignat}
\begin{alignat}{5}\label{lambda1}
\Lambda_{1} = \Lambda_{a 1} = \Lambda_{e 1} = \Lambda_{\tilde{\omega} 1} =
      \Lambda_{\sigma 1} = {} & - {} & {} & \left | \begin{array}{ccc}
      A_{\text{c}} & B_{\text{c}} & C_{\text{c}} \\
      G_{\text{c}} & H_{\text{c}} & I_{\text{c}} \\
      M_{\text{c}} & N_{\text{c}} & O_{\text{c}} \\
      \end{array} \right | {} & {} & - {} & {} & \left | \begin{array}{ccc}
      A_{\text{c}} & B_{\text{c}} & D_{\text{c}} \\
      G_{\text{c}} & H_{\text{c}} & J_{\text{c}} \\
      S_{\text{c}} & T_{\text{c}} & V_{\text{c}} \\
      \end{array} \right | {} & {} &
\notag \\
{} & - {} & {} & \left | \begin{array}{ccc}
      A_{\text{c}} & C_{\text{c}} & D_{\text{c}} \\
      M_{\text{c}} & O_{\text{c}} & P_{\text{c}} \\
      S_{\text{c}} & U_{\text{c}} & V_{\text{c}} \\
      \end{array} \right | {} & {} & - {} & {} & \left | \begin{array}{ccc}
      H_{\text{c}} & I_{\text{c}} & J_{\text{c}} \\
      N_{\text{c}} & O_{\text{c}} & P_{\text{c}} \\
      T_{\text{c}} & U_{\text{c}} & V_{\text{c}} \\
      \end{array} \right | ~, {} & {} &
\end{alignat}
\begin{align}\label{lambda0}
\Lambda_{0} = \Lambda_{a 0} = \Lambda_{e 0} = \Lambda_{\tilde{\omega} 0} =
      \Lambda_{\sigma 0} = \left | \begin{array}{cccc}
      A_{\text{c}} & B_{\text{c}} & C_{\text{c}} & D_{\text{c}} \\
      G_{\text{c}} & H_{\text{c}} & I_{\text{c}} & J_{\text{c}} \\
      M_{\text{c}} & N_{\text{c}} & O_{\text{c}} & P_{\text{c}} \\
      S_{\text{c}} & T_{\text{c}} & U_{\text{c}} & V_{\text{c}} \\
      \end{array} \right | ~,
\end{align}
\begin{alignat}{5}\label{lambdavt}
\Lambda_{a t} {} & = {} & {} & \left | \begin{array}{cccc}
      E_{\text{c}} & B_{\text{c}} & C_{\text{c}} & D_{\text{c}} \\
      K_{\text{c}} & H_{\text{c}} & I_{\text{c}} & J_{\text{c}} \\
      Q_{\text{c}} & N_{\text{c}} & O_{\text{c}} & P_{\text{c}} \\
      W_{\text{c}} & T_{\text{c}} & U_{\text{c}} & V_{\text{c}} \\
      \end{array} \right | {} & ~, ~~
\Lambda_{e t} {} & = {} & {} & \left | \begin{array}{cccc}
      A_{\text{c}} & E_{\text{c}} & C_{\text{c}} & D_{\text{c}} \\
      G_{\text{c}} & K_{\text{c}} & I_{\text{c}} & J_{\text{c}} \\
      M_{\text{c}} & Q_{\text{c}} & O_{\text{c}} & P_{\text{c}} \\
      S_{\text{c}} & W_{\text{c}} & U_{\text{c}} & V_{\text{c}} \\
      \end{array} \right | ~, {} & {} &
\notag \\
\Lambda_{\tilde{\omega} t} {} & = {} & {} & \left | \begin{array}{cccc}
      A_{\text{c}} & B_{\text{c}} & E_{\text{c}} & D_{\text{c}} \\
      G_{\text{c}} & H_{\text{c}} & K_{\text{c}} & J_{\text{c}} \\
      M_{\text{c}} & N_{\text{c}} & Q_{\text{c}} & P_{\text{c}} \\
      S_{\text{c}} & T_{\text{c}} & W_{\text{c}} & V_{\text{c}} \\
      \end{array} \right | {} & ~, ~~
\Lambda_{\sigma t} {} & = {} & {} & \left | \begin{array}{cccc}
      A_{\text{c}} & B_{\text{c}} & C_{\text{c}} & E_{\text{c}} \\
      G_{\text{c}} & H_{\text{c}} & I_{\text{c}} & K_{\text{c}} \\
      M_{\text{c}} & N_{\text{c}} & O_{\text{c}} & Q_{\text{c}} \\
      S_{\text{c}} & T_{\text{c}} & U_{\text{c}} & W_{\text{c}} \\
      \end{array} \right | ~. {} & {} &
\end{alignat}
\begin{alignat}{9}\label{lambdav}
\Lambda_{a} {} & = {} & {} & \left | \begin{array}{cccc}
      F_{\text{c}} & B_{\text{c}} & C_{\text{c}} & D_{\text{c}} \\
      L_{\text{c}} & H_{\text{c}} & I_{\text{c}} & J_{\text{c}} \\
      R_{\text{c}} & N_{\text{c}} & O_{\text{c}} & P_{\text{c}} \\
      X_{\text{c}} & T_{\text{c}} & U_{\text{c}} & V_{\text{c}} \\
      \end{array} \right | {} & {} & - {} & {} & \left | \begin{array}{ccc}
      E_{\text{c}} & B_{\text{c}} & C_{\text{c}} \\
      K_{\text{c}} & H_{\text{c}} & I_{\text{c}} \\
      Q_{\text{c}} & N_{\text{c}} & O_{\text{c}} \\
      \end{array} \right | {} & {} & - {} & {} & \left | \begin{array}{ccc}
      E_{\text{c}} & B_{\text{c}} & D_{\text{c}} \\
      K_{\text{c}} & H_{\text{c}} & J_{\text{c}} \\
      W_{\text{c}} & T_{\text{c}} & V_{\text{c}} \\
      \end{array} \right | {} & {} & - {} & {} & \left | \begin{array}{ccc}
      E_{\text{c}} & C_{\text{c}} & D_{\text{c}} \\
      Q_{\text{c}} & O_{\text{c}} & P_{\text{c}} \\
      W_{\text{c}} & U_{\text{c}} & V_{\text{c}} \\
      \end{array} \right | ~, {} & {} &
\notag \\
\Lambda_{e} {} & = {} & {} & \left | \begin{array}{cccc}
      A_{\text{c}} & F_{\text{c}} & C_{\text{c}} & D_{\text{c}} \\
      G_{\text{c}} & L_{\text{c}} & I_{\text{c}} & J_{\text{c}} \\
      M_{\text{c}} & R_{\text{c}} & O_{\text{c}} & P_{\text{c}} \\
      S_{\text{c}} & X_{\text{c}} & U_{\text{c}} & V_{\text{c}} \\
      \end{array} \right | {} & {} & - {} & {} & \left | \begin{array}{ccc}
      A_{\text{c}} & E_{\text{c}} & C_{\text{c}} \\
      G_{\text{c}} & K_{\text{c}} & I_{\text{c}} \\
      M_{\text{c}} & Q_{\text{c}} & O_{\text{c}} \\
      \end{array} \right | {} & {} & - {} & {} & \left | \begin{array}{ccc}
      A_{\text{c}} & E_{\text{c}} & D_{\text{c}} \\
      G_{\text{c}} & K_{\text{c}} & J_{\text{c}} \\
      S_{\text{c}} & W_{\text{c}} & V_{\text{c}} \\
      \end{array} \right | {} & {} & - {} & {} & \left | \begin{array}{ccc}
      K_{\text{c}} & I_{\text{c}} & J_{\text{c}} \\
      Q_{\text{c}} & O_{\text{c}} & P_{\text{c}} \\
      W_{\text{c}} & U_{\text{c}} & V_{\text{c}} \\
      \end{array} \right | ~, {} & {} &
\notag \\
\Lambda_{\tilde{\omega}} {} & = {} & {} & \left | \begin{array}{cccc}
      A_{\text{c}} & B_{\text{c}} & F_{\text{c}} & D_{\text{c}} \\
      G_{\text{c}} & H_{\text{c}} & L_{\text{c}} & J_{\text{c}} \\
      M_{\text{c}} & N_{\text{c}} & R_{\text{c}} & P_{\text{c}} \\
      S_{\text{c}} & T_{\text{c}} & X_{\text{c}} & V_{\text{c}} \\
      \end{array} \right | {} & {} & - {} & {} & \left | \begin{array}{ccc}
      A_{\text{c}} & B_{\text{c}} & E_{\text{c}} \\
      G_{\text{c}} & H_{\text{c}} & K_{\text{c}} \\
      M_{\text{c}} & N_{\text{c}} & Q_{\text{c}} \\
      \end{array} \right | {} & {} & - {} & {} & \left | \begin{array}{ccc}
      A_{\text{c}} & E_{\text{c}} & D_{\text{c}} \\
      M_{\text{c}} & Q_{\text{c}} & P_{\text{c}} \\
      S_{\text{c}} & W_{\text{c}} & V_{\text{c}} \\
      \end{array} \right | {} & {} & - {} & {} & \left | \begin{array}{ccc}
      H_{\text{c}} & K_{\text{c}} & J_{\text{c}} \\
      N_{\text{c}} & Q_{\text{c}} & P_{\text{c}} \\
      T_{\text{c}} & W_{\text{c}} & V_{\text{c}} \\
      \end{array} \right | ~, {} & {} &
\notag \\
\Lambda_{\sigma} {} & = {} & {} & \left | \begin{array}{cccc}
      A_{\text{c}} & B_{\text{c}} & C_{\text{c}} & F_{\text{c}} \\
      G_{\text{c}} & H_{\text{c}} & I_{\text{c}} & L_{\text{c}} \\
      M_{\text{c}} & N_{\text{c}} & O_{\text{c}} & R_{\text{c}} \\
      S_{\text{c}} & T_{\text{c}} & U_{\text{c}} & X_{\text{c}} \\
      \end{array} \right | {} & {} & - {} & {} & \left | \begin{array}{ccc}
      A_{\text{c}} & B_{\text{c}} & E_{\text{c}} \\
      G_{\text{c}} & H_{\text{c}} & K_{\text{c}} \\
      S_{\text{c}} & T_{\text{c}} & W_{\text{c}} \\
      \end{array} \right | {} & {} & - {} & {} & \left | \begin{array}{ccc}
      A_{\text{c}} & C_{\text{c}} & E_{\text{c}} \\
      M_{\text{c}} & O_{\text{c}} & Q_{\text{c}} \\
      S_{\text{c}} & U_{\text{c}} & W_{\text{c}} \\
      \end{array} \right | {} & {} & - {} & {} & \left | \begin{array}{ccc}
      H_{\text{c}} & I_{\text{c}} & K_{\text{c}} \\
      N_{\text{c}} & O_{\text{c}} & Q_{\text{c}} \\
      T_{\text{c}} & U_{\text{c}} & W_{\text{c}} \\
      \end{array} \right | ~. {} & {} &
\end{alignat}
The sought for solution of Eqs. (\ref{complete}) most significantly
depends on the fact whether the secular variations of the particle's orbit
caused by the non-gravitational effects depend on the orientation
of the orbit in space (the longitude of pericenter). After the averaging
over the synodic period the partial derivatives of the disturbing
function are not functions of the longitude of pericenter
\begin{equation}\label{pRi}
\frac{\partial}{\partial \tilde{\omega}}
\frac{\partial R}{\partial^{\star} a} =
\frac{\partial}{\partial \tilde{\omega}}
\frac{\partial R}{\partial e} =
\frac{\partial}{\partial \tilde{\omega}}
\frac{\partial R}{\partial \sigma} = 0 ~.
\end{equation}

\subsection{Linearization solution for non-gravitational effects with
rotational symmetry}
\label{subsec:solsymmetry}

In problems with the rotational symmetry around the star the terms
caused by the non-gravitational effects are not functions of the longitude
of pericenter \citep[see Appendix A in][]{kh}
\begin{equation}\label{pngi}
\frac{\partial}{\partial \tilde{\omega}}
\left ( \frac{da}{dt} \right )_{\text{EF}} =
\frac{\partial}{\partial \tilde{\omega}}
\left ( \frac{de}{dt} \right )_{\text{EF}} =
\frac{\partial}{\partial \tilde{\omega}}
\left ( \frac{d \tilde{\omega}}{dt} \right )_{\text{EF}} =
\frac{\partial}{\partial \tilde{\omega}}
\left ( \frac{d \sigma_{\text{b}}}{dt} +
t \frac{dn}{dt} \right )_{\text{EF}} = 0 ~.
\end{equation}
If we use properties shown in Eqs. (\ref{pRi}) and (\ref{pngi}) in
the calculation of the constants in Eqs. (\ref{oscillations}), then we
obtain for the non-gravitational effects with the rotational symmetry
\begin{equation}\label{zeros}
C_{\text{c}} = I_{\text{c}} = O_{\text{c}} = U_{\text{c}} = 0
\end{equation}
and the variations of $a$, $e$, and $\sigma$ are independent
of the variation of $\tilde{\omega}$ (see Eqs. \ref{oscillations}). In this
case the determinants $\Lambda_{0}$, $\Lambda_{a t}$, $\Lambda_{e t}$, and
$\Lambda_{\sigma t}$ in equations in Eqs. (\ref{lambda0})--(\ref{lambdavt})
give
\begin{equation}\label{reduction}
\Lambda_{0} = \Lambda_{a t} = \Lambda_{e t} = \Lambda_{\sigma t} = 0 ~.
\end{equation}
General solution of Eqs. (\ref{complete}) with substituted $\Lambda_{0}$ $=$ 0
has form
\begin{equation}\label{cubicsolution}
\delta_{\diamond} =
\frac{A_{\diamond 1}}{\lambda_{1}} \text{e}^{\lambda_{1} t} +
\frac{A_{\diamond 2}}{\lambda_{2}} \text{e}^{\lambda_{2} t} +
\frac{A_{\diamond 3}}{\lambda_{3}} \text{e}^{\lambda_{3} t} -
\frac{\Lambda_{\diamond t}}{2 \Lambda_{1}} t^{2} +
\frac{\Lambda_{2} \Lambda_{\diamond t} - \Lambda_{1} \Lambda_{\diamond}}
{\Lambda_{1}^{2}} t + B_{\diamond} ~,
\end{equation}
where subscript $\diamond$ represents one of the variables $a$,
$e$, $\tilde{\omega}$, or $\sigma$. $A_{\diamond i}$ are complex constants,
$B_{\diamond}$ are real constant numbers (as we will see later), and
$\lambda_{i}$ with $i$ $=$ 1, 2, 3 are all roots of the characteristic cubic
equation with real coefficients
\begin{equation}\label{cubic}
\lambda^{3} + \Lambda_{3} \lambda^{2} + \Lambda_{2} \lambda +
\Lambda_{1} = 0 ~.
\end{equation}
The roots of any cubic equation with real coefficients are always three
real numbers or one real number and two complex numbers that are
complex conjugate to each other. Equations (\ref{reduction}) and
(\ref{cubicsolution}) imply that the semimajor axis, eccentricity,
and resonant angular variable cannot have the terms varying quadratically
with the time for this linearized system. However, the longitude
of pericenter can have the term varying linearly with the time even in
the case when the partial derivatives with respect to the time
($E_{\text{c}}$, $K_{\text{c}}$, $Q_{\text{c}}$, $W_{\text{c}}$)
are zero (Eqs. \ref{lambdav}). The next step is the calculation
of the complex constants $A_{a i}$, $A_{e i}$, $A_{\tilde{\omega} i}$ and
$A_{\sigma i}$ as well as $B_{a}$, $B_{e}$, $B_{\tilde{\omega}}$ and
$B_{\sigma}$ from the initial conditions. This is standard procedure and
will be not shown here. The obtained equations for $A_{a i}$, $A_{e i}$,
$A_{\tilde{\omega} i}$ and $A_{\sigma i}$ are
\begin{align}\label{Ai}
A_{\diamond 1} &= \frac{\dddot{\delta}_{\diamond} (0) -
      \left ( \ddot{\delta}_{\diamond} (0) +
      \frac{\Lambda_{\diamond t}}{\Lambda_{1}} \right )
      \left ( \lambda_{2} + \lambda_{3} \right ) +
      \left ( \dot{\delta}_{\diamond} (0) -
      \frac{\Lambda_{2} \Lambda_{\diamond t} -
      \Lambda_{1} \Lambda_{\diamond}}{\Lambda_{1}^{2}} \right )
      \lambda_{2} \lambda_{3}}{\left ( \lambda_{1} - \lambda_{2} \right )
      \left ( \lambda_{1} - \lambda_{3} \right )} ~,
\notag \\
A_{\diamond 2} &= \frac{\dddot{\delta}_{\diamond} (0) -
      \left ( \ddot{\delta}_{\diamond} (0) +
      \frac{\Lambda_{\diamond t}}{\Lambda_{1}} \right )
      \left ( \lambda_{1} + \lambda_{3} \right ) +
      \left ( \dot{\delta}_{\diamond} (0) -
      \frac{\Lambda_{2} \Lambda_{\diamond t} -
      \Lambda_{1} \Lambda_{\diamond}}{\Lambda_{1}^{2}} \right )
      \lambda_{1} \lambda_{3}}{\left ( \lambda_{1} - \lambda_{2} \right )
      \left ( \lambda_{3} - \lambda_{2} \right )} ~,
\notag \\
A_{\diamond 3} &= \frac{\dddot{\delta}_{\diamond} (0) -
      \left ( \ddot{\delta}_{\diamond} (0) +
      \frac{\Lambda_{\diamond t}}{\Lambda_{1}} \right )
      \left ( \lambda_{1} + \lambda_{2} \right ) +
      \left ( \dot{\delta}_{\diamond} (0) -
      \frac{\Lambda_{2} \Lambda_{\diamond t} -
      \Lambda_{1} \Lambda_{\diamond}}{\Lambda_{1}^{2}} \right )
      \lambda_{1} \lambda_{2}}{\left ( \lambda_{1} - \lambda_{3} \right )
      \left ( \lambda_{2} - \lambda_{3} \right )} ~.
\end{align}
$A_{\diamond i}$ in Eqs. (\ref{Ai}) are related to $\lambda_{i}$ in such
a way that if $\lambda_{1}$ and $\lambda_{2}$ are complex conjugate to each
other and $\lambda_{3}$ is a real number, then also $A_{\diamond 1}$ and
$A_{\diamond 2}$ are complex conjugate to each other and $A_{\diamond 3}$ is
a real number. This property holds for any permutation of not equal
indexes $i$. Now, from $\delta_{\diamond} (0)$ $=$ 0 we can obtain $B_{a}$,
$B_{e}$, $B_{\tilde{\omega}}$ and $B_{\sigma}$ as follows
\begin{align}\label{Bdiamond}
B_{\diamond} = {} & - \Biggl [ \dddot{\delta}_{\diamond} (0) -
      \left ( \ddot{\delta}_{\diamond} (0) +
      \frac{\Lambda_{\diamond t}}{\Lambda_{1}} \right )
      \left ( \lambda_{1} + \lambda_{2} + \lambda_{3} \right )
\notag \\
& + \left ( \dot{\delta}_{\diamond} (0) -
      \frac{\Lambda_{2} \Lambda_{\diamond t} -
      \Lambda_{1} \Lambda_{\diamond}}{\Lambda_{1}^{2}} \right )
      \left ( \lambda_{1} \lambda_{2} + \lambda_{1} \lambda_{3} +
      \lambda_{2} \lambda_{3} \right ) \Biggr ] /
      \left ( \lambda_{1} \lambda_{2} \lambda_{3} \right ) ~.
\end{align}
$B_{\diamond}$ are always real numbers. It is interesting to note
that the linearization solution obtained for the case when the evolution
of longitude of pericenter is ignored in Eqs. (\ref{evolution}) is equivalent
with the solutions in Eq. (\ref{cubicsolution}) for $\delta_{a}$,
$\delta_{e}$, and $\delta_{\sigma}$ (Appendix \ref{app:equivalency}).

\subsection{Linearization solution for non-gravitational effects without
rotational symmetry}
\label{subsec:solasymmetry}

For the non-gravitational effects leading to the secular variations that
depend on the longitude of pericenter (in problems without the rotational
symmetry), the partial derivatives with respect to the longitude
of pericenter in Eqs. (\ref{linear}) are not equal to zero.
General solution of Eqs. (\ref{complete}) is in this case
\begin{equation}\label{quadricsolution}
\delta_{\diamond} =
C_{\diamond 1} \text{e}^{\lambda_{1} t} +
C_{\diamond 2} \text{e}^{\lambda_{2} t} +
C_{\diamond 3} \text{e}^{\lambda_{3} t} +
C_{\diamond 4} \text{e}^{\lambda_{4} t} -
\frac{\Lambda_{\diamond t}}{\Lambda_{0}} t +
\frac{\Lambda_{1} \Lambda_{\diamond t} - \Lambda_{0} \Lambda_{\diamond}}
{\Lambda_{0}^{2}} ~.
\end{equation}
Here $\lambda_{i}$ for $i$ $=$ 1, 2, 3, 4 are all roots of the characteristic
quadric equation with real coefficients
\begin{equation}\label{quadric}
\lambda^{4} + \Lambda_{3} \lambda^{3} + \Lambda_{2} \lambda^{2} +
\Lambda_{1} \lambda + \Lambda_{0} = 0 ~.
\end{equation}
The complex constants $C_{a i}$, $C_{e i}$, $C_{\tilde{\omega} i}$ and
$C_{\sigma i}$ can be determined from the initial conditions.
The obtained equations are
\begin{align}\label{Ci}
C_{\diamond 1} = {} & \Biggl [ \dddot{\delta}_{\diamond} (0) -
      \ddot{\delta}_{\diamond} (0) \left (
      \lambda_{2} + \lambda_{3} + \lambda_{4} \right ) +
      \left ( \dot{\delta}_{\diamond} (0) +
      \frac{\Lambda_{\diamond t}}{\Lambda_{0}} \right ) \left (
      \lambda_{2} \lambda_{3} +
      \lambda_{2} \lambda_{4} +
      \lambda_{3} \lambda_{4} \right )
\notag \\
& + \frac{\Lambda_{1} \Lambda_{\diamond t} - \Lambda_{0} \Lambda_{\diamond}}
      {\Lambda_{0}^{2}} \lambda_{2} \lambda_{3} \lambda_{4} \Biggr ] /
      \left [ \left ( \lambda_{1} - \lambda_{2} \right )
      \left ( \lambda_{1} - \lambda_{3} \right )
      \left ( \lambda_{1} - \lambda_{4} \right ) \right ] ~,
\notag \\
C_{\diamond 2} = {} & - \Biggl [ \dddot{\delta}_{\diamond} (0) -
      \ddot{\delta}_{\diamond} (0) \left (
      \lambda_{1} + \lambda_{3} + \lambda_{4} \right ) +
      \left ( \dot{\delta}_{\diamond} (0) +
      \frac{\Lambda_{\diamond t}}{\Lambda_{0}} \right ) \left (
      \lambda_{1} \lambda_{3} +
      \lambda_{1} \lambda_{4} +
      \lambda_{3} \lambda_{4} \right )
\notag \\
& + \frac{\Lambda_{1} \Lambda_{\diamond t} - \Lambda_{0} \Lambda_{\diamond}}
      {\Lambda_{0}^{2}} \lambda_{1} \lambda_{3} \lambda_{4} \Biggr ] /
      \left [ \left ( \lambda_{1} - \lambda_{2} \right )
      \left ( \lambda_{2} - \lambda_{3} \right )
      \left ( \lambda_{2} - \lambda_{4} \right ) \right ] ~,
\notag \\
C_{\diamond 3} = {} & \Biggl [ \dddot{\delta}_{\diamond} (0) -
      \ddot{\delta}_{\diamond} (0) \left (
      \lambda_{1} + \lambda_{2} + \lambda_{4} \right ) +
      \left ( \dot{\delta}_{\diamond} (0) +
      \frac{\Lambda_{\diamond t}}{\Lambda_{0}} \right ) \left (
      \lambda_{1} \lambda_{2} +
      \lambda_{1} \lambda_{4} +
      \lambda_{2} \lambda_{4} \right )
\notag \\
& + \frac{\Lambda_{1} \Lambda_{\diamond t} - \Lambda_{0} \Lambda_{\diamond}}
      {\Lambda_{0}^{2}} \lambda_{1} \lambda_{2} \lambda_{4} \Biggr ] /
      \left [ \left ( \lambda_{1} - \lambda_{3} \right )
      \left ( \lambda_{2} - \lambda_{3} \right )
      \left ( \lambda_{3} - \lambda_{4} \right ) \right ] ~,
\notag \\
C_{\diamond 4} = {} & - \Biggl [ \dddot{\delta}_{\diamond} (0) -
      \ddot{\delta}_{\diamond} (0) \left (
      \lambda_{1} + \lambda_{2} + \lambda_{3} \right ) +
      \left ( \dot{\delta}_{\diamond} (0) +
      \frac{\Lambda_{\diamond t}}{\Lambda_{0}} \right ) \left (
      \lambda_{1} \lambda_{2} +
      \lambda_{1} \lambda_{3} +
      \lambda_{2} \lambda_{3} \right )
\notag \\
& + \frac{\Lambda_{1} \Lambda_{\diamond t} - \Lambda_{0} \Lambda_{\diamond}}
      {\Lambda_{0}^{2}} \lambda_{1} \lambda_{2} \lambda_{3} \Biggr ] /
      \left [ \left ( \lambda_{1} - \lambda_{4} \right )
      \left ( \lambda_{2} - \lambda_{4} \right )
      \left ( \lambda_{3} - \lambda_{4} \right ) \right ] ~.
\end{align}
$C_{\diamond i}$ in Eqs. (\ref{Ci}) are related to $\lambda_{i}$ in such
a way that if $\lambda_{1}$ and $\lambda_{2}$ are real numbers and
$\lambda_{3}$ and $\lambda_{4}$ are complex conjugate to each other, then also
$C_{\diamond 1}$ and $C_{\diamond 2}$ are real numbers and $C_{\diamond 3}$
and $C_{\diamond 4}$ are complex conjugate to each other.
This property holds for any permutation of not equal indexes $i$.
For all $\lambda_{i}$ complex all $C_{\diamond i}$ are also complex
and the complex conjugacy is conserved.

\section{Stellar radiation as an example of non-gravitational effect
with rotational symmetry}
\label{sec:symmetry}

In this section the motion of a dust particle captured in a mean motion
resonance with a planet in a circular orbit around a radiating star will
be investigated. Secular variations of orbital parameters caused by
the stellar radiation will be used in order to verify the applicability
of the analytical approach derived in previous sections numerically.

\subsection{Equation of motion}
\label{subsec:speom}

Influence of electromagnetic radiation on the motion of a homogeneous
spherical dust particle can be described using the Poynting--Robertson (PR)
effect \citep{poynting,robertson,burns,klacka2004,icarus}. The acceleration
of the dust particle caused by the PR effect in a reference
frame associated with the source of radiation (star) is
\begin{equation}\label{PR}
\frac{d \vec{v}}{dt} = \beta \frac{\mu}{r^{2}}
\left [ \left ( 1 - \frac{\vec{v} \cdot \vec{e}_{\text{R}}}{c} \right )
\vec{e}_{\text{R}} - \frac{\vec{v}}{c} \right ] ~,
\end{equation}
where $r$ is the radial distance between the star and the dust particle,
$\vec{e}_{\text{R}}$ is the unit vector directed from the star
to the particle, $\vec{v}$ the velocity of the particle with respect
to the star, and $c$ is the speed of light in vacuum. The parameter $\beta$
is defined as the ratio between the electromagnetic radiation pressure
force and the gravitational force between the star and the particle
at rest with respect to the star
\begin{equation}\label{beta}
\beta = \frac{3 L_{\star} \bar{Q}'_{\text{pr}}}{16 \pi c \mu R_{\text{d}}
\varrho} ~,
\end{equation}
here $L_{\star}$ is the stellar luminosity, $\bar{Q}'_{\text{pr}}$ is
the dimensionless efficiency factor for the radiation pressure averaged
over the stellar spectrum and calculated for the radial direction
($\bar{Q}'_{\text{pr}}$ $=$ 1 for a perfectly absorbing sphere),
and $R_{\text{d}}$ is the radius of the dust particle with the mass density
$\varrho$.

Expanding solar corona supplies the observed continuous flux of the solar
wind inside the heliosphere formed by supersonic shock of the solar wind in
ambient moving interstellar matter. The interaction of stellar winds with
the interstellar matter has been directly observed at many stars.
The stellar wind can affect the motion of the dust particles
orbiting the star. It is possible to derive an acceleration
caused by wind corpuscules impinging on the dust particle using
a relativistic approach \citep{klasan,covsw}. For a radial stellar wind
the following acceleration affecting the dynamics of dust particles in
the accuracy to first order in $v / c$ ($v$ is the speed of the dust particle
with respect to the star), first order in $u / c$ ($u$ is the speed
of the stellar wind with respect to the star) and first order in $v / u$
can be derived
\begin{equation}\label{SW}
\frac{d \vec{v}}{dt} = \frac{\eta}{\bar{Q}'_{\text{pr}}}
\beta \frac{u}{c} \frac{\mu}{r^{2}} \left [
\left ( 1 - \frac{\vec{v} \cdot \vec{e}_{\text{R}}}{u} \right )
\vec{e}_{\text{R}} - \frac{\vec{v}}{u} \right ] ~.
\end{equation}
$\eta$ is to the given accuracy the ratio of the stellar wind energy to
the stellar electromagnetic radiation energy, both radiated per unit time
\begin{equation}\label{eta}
\eta = \frac{4 \pi r^{2} u}{L_{\star}}
\sum_{j = 1}^{N} n_{\text{sw}j} m_{\text{sw}j} c^{2} ~,
\end{equation}
where $m_{\text{sw}j}$ and $n_{\text{sw}j}$, $j$ $=$ 1 to $N$, are
the masses and concentrations of the stellar wind particles at a distance
$r$ from the star ($u$ $=$ 450 km/s and $\eta$ $=$ 0.38 for the Sun,
\citealt{covsw}).

When we add the gravitational accelerations from the star and the planet,
we obtain the final equation of motion of the dust grain in the PCRTBP
with the electromagnetic radiation and the radial stellar wind in the frame
of reference associated with the star
\begin{align}\label{speom}
\frac{d \vec{v}}{dt} = {} & - \frac{\mu}{r^{2}}
      \left ( 1 - \beta \right ) \vec{e}_{\text{R}} -
      \frac{G_{0} M_{\text{P}}}{\vert \vec{r} - \vec{r}_{\text{P}} \vert^{3}}
      (\vec{r} - \vec{r}_{\text{P}}) -
      \frac{G_{0} M_{\text{P}}}{r_{\text{P}}^{3}} \vec{r}_{\text{P}}
\notag \\
& - \beta \frac{\mu}{r^{2}}
      \left ( 1 + \frac{\eta}{\bar{Q}'_{\text{pr}}} \right )
      \left ( \frac{\vec{v} \cdot \vec{e}_{\text{R}}}{c}
      \vec{e}_{\text{R}} + \frac{\vec{v}}{c} \right ) ~.
\end{align}
In the equation above we have used the assumption that
$( \eta / \bar{Q}'_{\text{pr}} ) ( u / c )$ $\ll$ 1 at summation
of Eq. (\ref{PR}) and Eq. (\ref{SW}). The radial term not depending on
the particle's velocity in the PR effect can be added to the stellar gravity.

\subsection{Secular variations}
\label{subsec:spsec}

The acceleration caused by the PR effect and the radial stellar wind
in Eq. (\ref{speom}) can be used as the perturbation acceleration in
Gauss's perturbation equations of celestial mechanics \citep[e.g.][]{danby}.
The acceleration including all terms can be used as the perturbation.
But if we want to describe the motion of the cosmic dust particle
with slowly varying orbits, then it is convenient to use last
term in Eq. (\ref{speom}) as a perturbation to the orbital motion
in the gravitational field of a star with the reduced mass
$M_{\star} ( 1 - \beta )$. Gauss's perturbation equations then give
the following averaged values in Eqs. (\ref{evolution})
\begin{align}\label{spsecular}
\left ( \frac{da}{dt} \right )_{\text{EF}} = {} & -
      \frac{\beta \mu}{c a \alpha^{3}}
      \left ( 1 + \frac{\eta}{\bar{Q}'_{\text{pr}}} \right )
      \left ( 2 + 3 e^{2} \right ) ~,
\notag \\
\left ( \frac{de}{dt} \right )_{\text{EF}} = {} & -
      \frac{\beta \mu}{2 c a^{2} \alpha}
      \left ( 1 + \frac{\eta}{\bar{Q}'_{\text{pr}}} \right ) 5 e ~,
\notag \\
\left ( \frac{d \tilde{\omega}}{dt} \right )_{\text{EF}} = {} &
      0 ~,
\notag \\
\left ( \frac{d \sigma_{\text{b}}}{dt} +
      t \frac{dn}{dt} \right )_{\text{EF}} = {} & 0 ~.
\end{align}
In this case the expressions in the previous sections which
contain $L$ and $n$ must be modified. The modifying equations are
$L$ $=$ $\sqrt{\mu ( 1 - \beta ) a}$ and
$n$ $=$ $\sqrt{\mu ( 1 - \beta ) / a^{3}}$.

\subsection{Linearization of averaged resonant equations}
\label{subsec:splinearization}

The linearization of the averaged resonant equations (Eqs. \ref{evolution})
in neighborhood of the initial conditions (the semimajor axis,
eccentricity, longitude of pericenter, and resonant angular variable)
requires the knowledge of properties assigned by the averaging
to the partial derivatives of the disturbing function
(e.g. Eqs. \ref{relations} and Eqs. \ref{pRi}). The averaging
of the expressions containing the disturbing function $R$
can be done numerically for a mutual configuration of orbits
of the planet and the dust particle given by the initial
conditions from time zero to time equal to the synodic
period. The partial derivatives with respect to the time
in Eqs. (\ref{linear}) are zero in the PCRTBP with radiation since
none external variation is influencing the solved problem.
The partial derivatives with respect to $a$, $e$, $\tilde{\omega}$ and
$\sigma$ can be calculated using Eqs. (\ref{spsecular}) (see Eqs.
\ref{alphabet} without the terms from the interstellar gas flow in Appendix
\ref{app:coefficients}). The assumed rotational symmetry of the stellar
radiation gives $\Lambda_{0}$ $=$ 0 (Eqs. \ref{zeros}). The $\Lambda_{3}$,
$\Lambda_{2}$, and $\Lambda_{1}$ given by
Eqs. (\ref{lambda3})--(\ref{lambda1}) determine $\lambda_{i}$ as roots
of the cubic equation Eq. (\ref{cubic}). The oscillations are present in
a solution with one real root and two complex roots that are complex
conjugate to each other. The opposite imaginary parts of the two complex
$\lambda_{i}$ determine an angular frequency of the oscillation.

\subsection{Numerical checking}
\label{subsec:spcheck}

We are interested in an applicability of the linearization solution
derived analytically in Sect. \ref{sec:linearization} for librations in
the PCRTBP with radiation. The equation of motion (Eq. \ref{speom})
was solved numerically in order to determine a reference standard
for the libration in the mean motion resonances comparable with
the analytically derived linearization solution. Equations
(\ref{evolution}) are averaged over the synodic period. All initial
parameters in Eqs. (\ref{evolution}) obtained from the numerical solution
of the equation of motion were averaged over the first synodic period.
If we consider all $\beta$ for a given mean motion resonance in
the PCRTBP with radiation, then a phase space containing all
possible evolutions has four dimensions ($\beta$, $a$, $e$,
$\sigma$). The evolution of longitude of pericenter can be studied
separately (see Eqs. \ref{zeros}). For the sake of simplicity we will
vary only the eccentricity and the resonant angular variable in the initial
un-averaged phase space at fixed $\beta$ and a shift of the semimajor axis
from an exact resonance. The semimajor axis of the exact resonance will
be defined as $a_{\text{r}}$ $=$ $a_{\text{P}}$ $( 1 - \beta )^{1/3}$
$[ M_{\star} / ( M_{\star} + M_{\text{P}}) ]^{1/3}$ $[ p / ( p + q ) ]^{2/3}$.
From this definition we have for the shift $\Delta$ $=$ $a$ $-$ $a_{\text{r}}$.

The mean motion resonances can occur if the variation of the semimajor axis
caused by the non-gravitational effects can be compensated by the gravitational
influence of the planet. This implies the resonant condition in the PCRTBP
with radiation (see the first equation in Eqs. \ref{evolution})
\begin{equation}\label{spaxis}
\frac{da}{dt} = - \frac{2 s a}{L}
\frac{\partial R}{\partial \sigma} -
\frac{\beta \mu}{c a \alpha^{3}}
\left ( 1 + \frac{\eta}{\bar{Q}'_{\text{pr}}} \right )
\left ( 2 + 3 e^{2} \right ) = 0 ~.
\end{equation}
Form the equation above the resonant angular variable of the particle
for some shift and some eccentricity can be calculated. In a conservative
PCRTBP the $kh$ plane defined by $k$ $=$ $e \cos \sigma$ and $h$ $=$
$e \sin \sigma$ is commonly used for exploring properties
belonging to the Hamiltonian in the mean motion resonances
\citep[e.g.][]{greenberg,beauge,mude}. $k$ and $h$ are the non-canonical
variables used also in non-conservative cases \citep{FM1,FM2,sidnes}.

\begin{figure}[t]
\begin{center}
\includegraphics[width=0.83503521126760563380281690140845\textwidth]{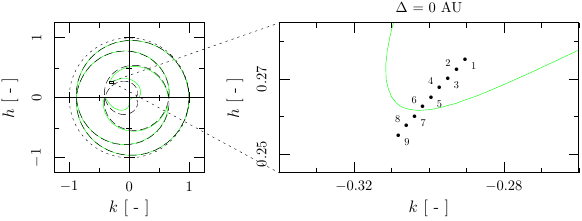}
\end{center}
\caption{The left-hand side plot depicts in the $kh$ plane ($k$ $=$
$e \cos \sigma$ and $h$ $=$ $e \sin \sigma$) the solutions of resonant
condition in the planar circular restricted Sun-Earth-dust problem with
solar radiation (green solid line) for a particle with $R$ $=$ 10 $\mu$m,
$\varrho$ $=$ 2 g.cm$^{-3}$, and $\bar{Q}'_{\text{pr}}$ $=$ 1 in
the exterior mean motion 6/5 orbital resonance obtained for zero shift.
Collisions of the planet and the particle can occur during the synodic
period in the shown locations (black dashed line). The right-hand side
plot shows locations of initial $kh$ points belonging to evolutions
depicted in Figs. \ref{fig:spap} and \ref{fig:spshift} (circles).}
\label{fig:spkh}
\end{figure}

\begin{figure}[t]
\begin{center}
\includegraphics[width=0.85\textwidth]{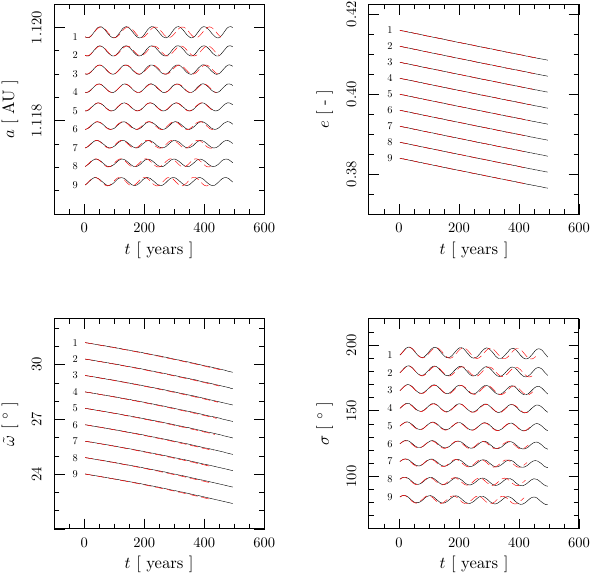}
\end{center}
\caption{Evolutions of the semimajor axis, eccentricity, longitude
of perihelion, and resonant angular variable averaged over the synodic
period in the planar circular restricted Sun-Earth-dust problem with
the solar radiation. A dust particle with $R$ $=$ 10 $\mu$m, $\varrho$
$=$ 2 g.cm$^{-3}$, and $\bar{Q}'_{\text{pr}}$ $=$ 1 is initially located
in the exterior mean motion 6/5 resonance. The numerical solutions of equation
of motion (Eq. \ref{speom}, black solid line) are compared with
the linearization solutions (Eq. \ref{cubicsolution}, red dashed line).
The evolutions starting with various $\tilde{\omega}$ and $\sigma$
(see text) are successively translated by 4 $\times$ 10$^{-4}$ AU, 4 $\times$
10$^{-3}$, 1$^{\circ}$, and 14$^{\circ}$ in $a$, $e$, $\tilde{\omega}$ and
$\sigma$, respectively. The evolution with the number of initial averaged
conditions 5 is not translated.}
\label{fig:spap}
\end{figure}

The averaged phase space reduced to the eccentricity and the resonant
angular variable is depicted in the left-hand side plot of Fig. \ref{fig:spkh}
as the $kh$ plane. The green solid line in Fig. \ref{fig:spkh} shows numerical
solutions of the resonant condition at the shift equal to zero for a particle
with $R$ $=$ 10 $\mu$m, $\varrho$ $=$ 2 g.cm$^{-3}$, and $\bar{Q}'_{\text{pr}}$
$=$ 1 in the exterior mean motion 6/5 orbital resonance with the Earth.
The purpose of the left-hand side plot in Fig. \ref{fig:spkh} is to show
locations where the captures into the resonance are possible. The shift
zero is used here in order to show later how a non-zero shift influences
the solutions of resonant condition. In the $kh$ plane the collisions
of the planet with the particle occur on the black dashed curve in
Fig. \ref{fig:spkh}. The dashed curves cannot be crossed by the $kh$ point
during the evolution in a mean motion resonance. The black rectangle in
the left-hand side plot of Fig. \ref{fig:spkh} contains initial $kh$ points
for the evolutions depicted in Figs. \ref{fig:spap} and \ref{fig:spshift}.
The initial $kh$ points are in the averaged phase space calculated using
evolving $e$ and $\sigma$ averaged over the first synodic period
of the corresponding evolutions.

\begin{figure}[t]
\begin{center}
\includegraphics[width=0.83503521126760563380281690140845\textwidth]{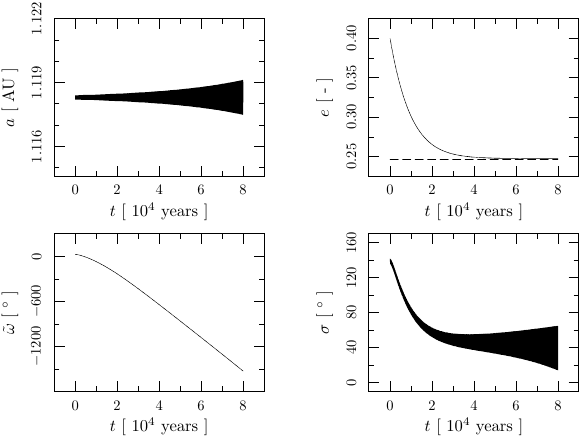}
\end{center}
\caption{The numerical solution with the number 5 from Fig. \ref{fig:spap}
integrated over an interval 8 $\times$ 10$^{4}$ years (black solid line).
The interval ends with an asymptotic approach of the eccentricity librations
to the universal eccentricity. The horizontal line in the eccentricity plot
depicts the universal eccentricity calculated from Eq. (\ref{universal}) for
the exterior 6/5 resonance $e_{\text{u}}$ $\approx$ 0.2472 (black dashed
line).}
\label{fig:unie}
\end{figure}

A comparison of the numerical (Eq. \ref{speom}, black solid line) and
the analytical (Eq. \ref{cubicsolution}, red dashed line) solution
is depicted in Fig. \ref{fig:spap}. Initial conditions for oscular
parameters are $\Delta_{\text{in}}$ $=$ 0 AU, $e_{\text{in}}$ $=$ 0.4,
$\tilde{\omega}$ $=$ $-$ $\sigma_{\text{in}}$ $q / ( p + q ) $,
and $\sigma_{\text{in}}$ $\in$ $\{ 136^{\circ}, 136.5^{\circ},
137^{\circ}, ..., 140^{\circ} \}$.
Initial true anomalies of the planet and the particle were zero.
The curves for $a$, $e$, $\tilde{\omega}$ and $\sigma$ obtained using
the various initial conditions are successive translated by 4 $\times$
10$^{-4}$ AU, 4 $\times$ 10$^{-3}$, 1$^{\circ}$, and 14$^{\circ}$,
respectively. Zero translation is at the evolution with the number
of initial averaged conditions 5. Without the translation the evolutions
of $a$, $e$, and $\sigma$ would be overlapped. The evolutions
of $\tilde{\omega}$ without the translation would be shown in the opposite
order with a small separation.

The variations of the frequency and the libration amplitude with the initial
conditions can be easily seen in the evolutions of the semimajor axis and
the resonant angular variable. The evolution of eccentricity is determined
by the second equation in Eqs. (\ref{evolution}). By solving the condition
$de / dt$ $=$ 0 at the solution of resonant condition ($da / dt$ $=$ 0) we
obtain the so called ``universal eccentricity'' \citep{FM2}. The universal
eccentricity in the PCRTBP with radiation exists only for the exterior
resonances. For the universal eccentricity ($e_{\text{u}}$) holds
\begin{equation}\label{universal}
1 - \frac{3 e_{\text{u}}^{2} + 2}
{2 \left ( 1 - e_{\text{u}}^{2} \right )^{3/2}} \frac{p + q}{p} = 0 ~.
\end{equation}
The universal eccentricity in the exterior resonance (given by
the resonant numbers $p$ and $q$) is equal for all particles.
By the averaging of the second equation in Eqs. (\ref{evolution}) over
the libration period one obtains governing differential equation
for the evolution of eccentricity \citep{LZ1997}. The disturbing
function in the equation can be hidden using identity $da / dt$ $=$ 0
that is valid for the resonances after the averaging over the libration
period. The solution of the condition $de / dt$ $=$ 0 for
the exterior resonances is the universal eccentricity. Hence,
the eccentricity is constant at the universal eccentricity after
the averaging over the libration period. If the initial eccentricity
averaged over the libration period is not equal to the universal
eccentricity, then the eccentricity in the exterior resonances asymptotically
approaches the universal eccentricity. For the exterior 6/5 resonance
$e_{\text{u}}$ $\approx$ 0.2472 and the evolutions of eccentricity in
Fig. \ref{fig:spap} decrease to this value. But the asymptotic value is yet
distant from the eccentricities in Fig. \ref{fig:spap}. In order to
show the slowed-down approach of the eccentricity to the universal
eccentricity for the evolution number 5 from Fig. \ref{fig:spap} we
integrated the evolution over an interval of 8 $\times$ 10$^{4}$ years
in Fig. \ref{fig:unie}. At the ends of $\tilde{\omega}$ evolutions
in Fig. \ref{fig:spap} the linearization solution gives systematically
smaller $\tilde{\omega}$ than the solution of the equation of motion.
These differences were found to be dependent on $\tilde{\omega}$ used
in the averaging of the terms with the disturbing function. For a different
$\tilde{\omega}$ giving the same $\sigma$ \citep[see Fig. 6 in][]{kh}
the linearization solution can give also slightly larger $\tilde{\omega}$
at the ends of the evolutions. Therefore, this should be numerical feature
since in the reality the dependence on $\tilde{\omega}$ should not exists.
The libration of $\sigma$ occurs close to $\sigma$ satisfying the resonant
condition (Eq. \ref{spaxis}).

\begin{figure}[t]
\begin{center}
\includegraphics[width=0.90088028169014084507042253521127\textwidth]{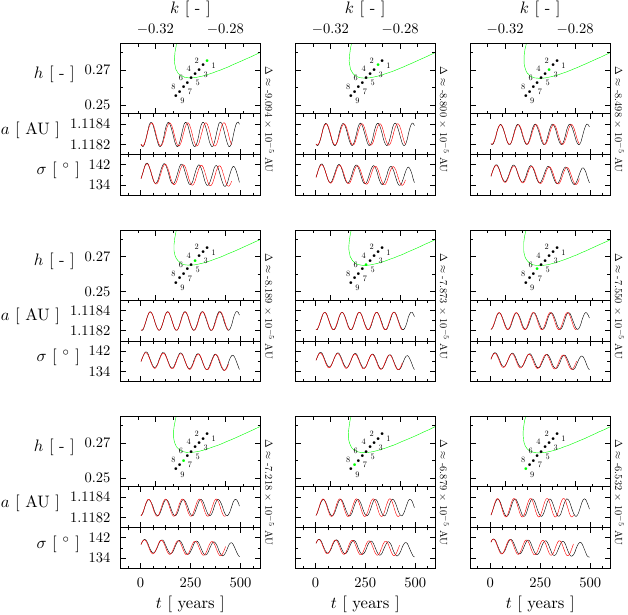}
\end{center}
\caption{The $kh$ points calculated from the initial averaged $e$ and
$\sigma$ of the evolutions in Fig. \ref{fig:spap} are shown in the top
panel of each plot (circles). The green $kh$ point marks the evolution with
the initial shift in the averaged phase space ($\Delta$) written on
the right-hand side of plot. The solutions of resonant condition (green
solid lines) are obtained for the initial averaged shift. The evolutions
of semimajor axis obtained numerically (black solid line) and analytically
(red solid line) are compared in the middle panel of each plot. The bottom
panels show the compared evolutions of the resonant angular variable.}
\label{fig:spshift}
\end{figure}

Main purpose of nine plots in Fig. \ref{fig:spshift} is to compare
variations of the solutions of resonant condition with the varying
initial conditions of the evolutions in Fig. \ref{fig:spap} in the averaged
phase space. The top panel of each plot shows the part of the $kh$ plane
equal to the black rectangle in the left-hand side plot of Fig. \ref{fig:spkh}.
The initial $kh$ points calculated using $e$ and $\sigma$ averaged over
the first synodic period are also shown (circles). The green circle
denotes the $kh$ point of the evolution that has the initial averaged
shift shown on the right-hand side of each plot. In the averaged
phase space the shift differs from the initial shift (zero) used in
the un-averaged phase space. The differences exist also for the eccentricity
and the resonant angular variable. The shown part of the $kh$ plane
depicts the eccentricities and the resonant angular variables satisfying
the resonant condition at the initial averaged shift (green solid line).
The variations of the positions of green solid lines due to the varying
initial averaged shift are smaller than the variations of the initial
$kh$ points. This can be seen in Fig. \ref{fig:spshift} if we compare
the solutions of resonant condition at the evolutions 1 and 9. The green
solid line at the evolution 1 is between the $kh$ points 4 and 5 and
at the evolution 9 is between the $kh$ points 5 and 6. The green solid
line obtained at the zero shift in the right-hand side plot
of Fig. \ref{fig:spkh} is between the $kh$ points 6 and 7. It is not easy
to hit the solution of resonant condition with the $kh$ point since
the averaged values obtained from the numerical solution are discrete.

\begin{figure}[t]
\begin{center}
\includegraphics[width=0.85\textwidth]{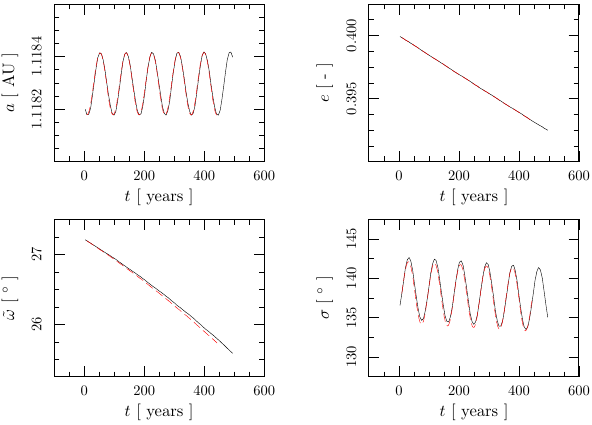}
\end{center}
\caption{The evolution 1 from Fig. \ref{fig:spap} (black solid
line) approximated by the linearization solution determined in the averaged
phase space with orbital parameters from the first minimum
of the semimajor axis (red dashed line).}
\label{fig:spcom}
\end{figure}

The middle panel of Fig. \ref{fig:spshift} shows compared evolutions
of the semimajor axis obtained numerically (black solid line) and
analytically (red solid line) that are shown and numbered also in
Fig. \ref{fig:spap}. Similarly the bottom panel shows the compared
evolutions of the resonant angular variable. The best frequency
accordance between the analytical and the numerical solution is found
at the evolution 5. The initial conditions in the averaged
phase space of the evolution 5 are closest to the solution
of the resonant condition at the used initial averaged shift
(see Fig. \ref{fig:spshift}). In some cases (mentioned later) the initial
conditions closest to the solution of the resonant condition do not
give the best accordance between the analytical and the numerical
solution. But the best accordance is usually found for the initial
conditions not far from this solution.

The property that the best frequency accordance is found at
the solution of resonant condition is caused by the evolution
of the resonant angular variable. The frequency calculated from
the linearization solution in the considered problem most sensitively
depends on the initial averaged value of $\sigma$. The dependencies on other
initial averaged orbital parameters are much smaller. The linearization
frequency is determined in such a way that the evolution of the resonant
angular variable is best approximated during a short time interval after
the initial time. The found linearization frequency does not have to describe
the real libration frequency but the evolutions of orbital parameters have
to be correctly described during a sort time interval after the initial
time. The linearization frequency significantly varies during librations
of $\sigma$ in Fig. \ref{fig:spap}.

At the solution of resonant condition the averaged time derivative
of the semimajor axis is zero. For the resonant angular variable we have
\begin{equation}\label{avesigma}
\sigma = \frac{p + q}{q} \lambda_{\text{P}} - s \lambda - \tilde{\omega} =
\left ( \frac{p + q}{q} n_{\text{P}} - s n \right ) t +
\frac{p + q}{q} \lambda_{\text{P}0} - s (\sigma_{\text{b}} +
\tilde{\omega}) - \tilde{\omega} ~.
\end{equation}
$n$, $\sigma_{\text{b}}$, and $\tilde{\omega}$ are constant and $\sigma$
depends linearly on time in Keplerian approximation of the motion
during the synodic period. When $da / dt$ $=$ 0 in the numerically
averaged evolution, then the shift from the exact resonance is minimal or
maximal. In the considered approximation the linear time dependence
of $\sigma$ is steepest at the solution of resonant condition. The solution
of resonant condition is approximately in the middle of $\sigma$ libration.
The resonant libration frequency is more accurately determined using
the linearization solution when the initial averaged conditions are
closer to the solution of resonant condition. The entire evolution during
more librations is in this case sufficiently well approximated.

Even for the evolution 1 in Fig. \ref{fig:spap} we can obtain an usable
linearization solution if we use the initial averaged conditions
from later time that are close to the solution of resonant condition
for the calculation of the parameters of linearization solution. In
other words, the initial averaged conditions close to the minimum
or maximum of the semimajor axis. The numerical integration can also
start with positions and velocities from later time that are close to
the minimum or maximum of the semimajor axis in order to obtain the usable
initial averaged conditions in the first synodic period. Such a case
is depicted in Fig. \ref{fig:spcom}. However, we must note that
the real increase of the semimajor axis during the first libration in
Fig. \ref{fig:spcom} is faster then the increase predicted by
the linearization solution and the decrease is slower. This holds also
for the evolution 5 in Fig. \ref{fig:spap}.

Figs. \ref{fig:spap}, \ref{fig:spshift}, and \ref{fig:spcom} show
the applicability of the linearization solution for the exterior 6/5
resonance with the Earth at the vicinity of one $\sigma$ satisfying
the resonant condition at the eccentricity 0.4. The applicability
of the linearization solution was checked for various exterior
resonances at the vicinity of $\sigma$ satisfying the resonant condition
for the eccentricities up to $\sim$0.6.

\begin{figure}[t]
\begin{center}
\includegraphics[width=0.85\textwidth]{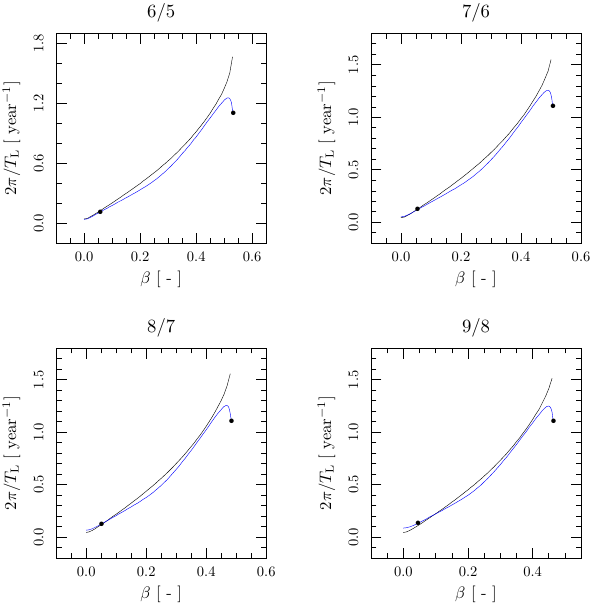}
\end{center}
\caption{Angular frequencies of the libration for dust particles with various
$\beta$ after an infinitesimal displacement from the periodic solutions
calculated for 6/5, 7/6, 8/7, and 9/8 exterior resonances in the planar
circular restricted Sun-Earth-dust problem with the PR effect and the radial
solar wind. The same color in the plots denotes the sets of the periodic
solutions emerging from the same $\sigma$ as $\beta$ increases from zero.
Circles depict the frequencies of libration for the periodic solutions in
the exact resonances \citep{kh}.}
\label{fig:ref}
\end{figure}

The resonant angular variable is commonly determined using the initial
conditions from the un-averaged phase space with the zero shift.
The linearization frequency found for the evolution starting with these
initial parameters can be different from the real libration frequency
of the evolution. Main reason is the fact that the evolution starting
with zero shift in the un-averaged phase space usually does not start
at the minimal or maximal semimajor axis in the averaged phase space
(in the solution of resonant condition). Although the zero initial shift
in the un-averaged phase space usually gives the evolution with a non-zero
initial shift in the averaged phase space.

The statements in the previous paragraph can be easily verified
if we compare the right-hand side plot in Figs. \ref{fig:spkh} with the plots
in Fig. \ref{fig:spshift}. The best frequency accordance is obtained for
the evolution 5 that has the non-zero initial shift from
the exact resonance in the averaged phase space. The resonant angular
variable obtained from the solution of resonant condition at the zero
shift in the right-hand side plot of Fig. \ref{fig:spkh} is between
the initial averaged conditions of the evolutions 6 and 7.
The linearization frequency found for $a$ and $e$ from
the un-averaged phase space and this $\sigma$ would be between
the linearization frequencies found for the evolutions 6 and 7.
The evolutions 6 and 7 in Fig. \ref{fig:spshift} have the zero
initial shift in the un-averaged phase space and do not give the best
frequency accordance.

The parameters giving the linearization solution for the evolution 5
in Fig. \ref{fig:spap} are shown in Appendix \ref{app:lp}
(Table \ref{tab:splp}). Since the real parts of $\lambda_{1}$ and
$\lambda_{3}$ in Table \ref{tab:splp} are positive the libration amplitude
of the linearization solution increases. This is usually interpreted as
an instability. The capture with the non-zero libration amplitude in
the PCRTBP with radiation is only temporary. The non-zero libration amplitude
increases in accordance with the results valid for the PR effect in
\citet{gomes95}.

\subsection{Periodic solutions}
\label{subsec:spperiod}

For the exterior resonances in the PCRTBP with radiation periodic solutions
exist. These periodic solutions set maximal capture time in the exterior
resonances theoretically to infinity. Their position in $ae\sigma$ phase
space can be obtained as points where $a$, $e$, and $\sigma$ are
constant after the averaging over the synodic period for a particle
with given $\beta$ \citep{kh}. The periodic solutions exist at
the universal eccentricity. The libration amplitude of the periodical
solutions is zero \citep{kh}. Using analytical theory from \citet{gomes95}
can be proved that the zero libration amplitude does not increase in contrary
to the cases with the non-zero libration amplitude. From a theoretical
point of view the libration is consistent with a ``libration''
of a pendulum in an equilibrium point. The frequency of libration can
be defined also for these periodic solutions as a frequency of the libration
after an infinitesimal displacement from the periodic solution. Such
frequencies are calculated in Fig. \ref{fig:ref} for the periodic solutions
in 6/5, 7/6, 8/7, and 9/8 exterior resonances with the Earth in a circular
orbit around radiating Sun. Since the periodic solutions exist at the solution
of resonant condition ($da / dt$ $=$ 0) the frequencies are correctly
determined from the linearization theory. For periodic solutions
the linearization solution does not give exactly the zero libration
amplitude, but the obtained libration amplitude is very small.

In the conservative PCRTBP the periodic solutions in the mean motion
resonances exist at various eccentricities. These periodic solutions can be
obtained using the method in \citet{kh} without the condition giving
the universal eccentricity. Periodic solutions in the circular-planar,
spatial-circular, elliptic-planar and spatial-elliptic restricted three-body
problem with the PR effect were found to exist for the dust particles captured
in the mean motion 1/1 resonance with the planet \citep{points,triangular}.

\section{Interstellar gas flow as an example of non-gravitational effect
without rotational symmetry}
\label{sec:asymmetry}

Non-gravitational effects secularly varying orbits in a dependence on
their orientation in space are not often considered in
the literature. An interstellar gas entering an astrosphere
of the star varies the orbits in such a way. The secular
variation of orbit in this case depends on the orientation of orbit
with respect to an interstellar gas velocity vector. In this section
we use secular variations of orbital parameters caused by the stellar
radiation and the interstellar gas flow to verify the applicability
of the analytical approach derived in Sect. \ref{sec:linearization}.

\subsection{Equation of motion}
\label{subsec:mneom}

The interstellar matter containing $i$ gas components with temperatures
$T_{i}$ moving with a relative velocity $\vec{v_{\text{F}}}$ with respect
to the star affects the dynamic of a spherical dust particle according to
\citet{baines} with the acceleration
\begin{equation}\label{ISW}
\frac{d \vec{v}}{dt} = - \sum_{i = 1}^{N} c_{\text{D}i} \gamma_{i}
\vert \vec{v} - \vec{v}_{\text{F}} \vert
\left ( \vec{v} - \vec{v}_{\text{F}} \right ) ~.
\end{equation}
$c_{\text{D}i}$ in Eq. (\ref{ISW}) is the drag
coefficient
\begin{align}\label{cd}
c_{\text{D}i}(s_{i}) = {} & \frac{1}{\sqrt{\pi}}
      \left ( \frac{1}{s_{i}} + \frac{1}{2 s_{i}^{3}} \right )
      \text{e}^{-s_{i}^{2}} +
      \left ( 1 + \frac{1}{s_{i}^{2}} - \frac{1}{4 s_{i}^{4}} \right )
      \text{erf}(s_{i})
\notag \\
& + \left ( 1 - \delta_{i} \right )
      \left ( \frac{T_{\text{d}}}{T_{i}} \right )^{1 / 2}
      \frac{\sqrt{\pi}}{3s_{i}} ~,
\end{align}
where erf$(s_{i})$ is the error function $\text{erf}(s_{i})$ $=$
$2 / \sqrt{\pi} \int_{0}^{s_{i}} \text{e}^{-t_{\text{p}}^{2}} dt_{\text{p}}$,
$\delta_{i}$ is the fraction of impinging particles specularly reflected at
the surface (a diffuse reflection is assumed for the rest of the particles, see
\citealt{baines,gustafson}), $T_{\text{d}}$ is the temperature of the dust
grain. $s_{i}$ in Eq. (\ref{cd}) is the molecular speed ratio
\begin{equation}\label{s}
s_{i} = \sqrt{\frac{m_{i}}{2 k T_{i}}} U ~.
\end{equation}
Here $m_{i}$ is the mass of the atom in the $i$th gas component,
$k$ is Boltzmann's constant, and
$U$ $=$ $\vert \vec{v} - \vec{v}_{\text{F}} \vert$
is the relative speed of the dust particle with respect to the gas.
For the collision parameter $\gamma_{i}$ in Eq. (\ref{ISW}) we find
\begin{equation}\label{cp}
\gamma_{i} = n_{i} \frac{m_{i}}{m} A' ~,
\end{equation}
where $n_{i}$ is the number density of the $i$th gas component,
and $A'$ is the geometrical cross section of the dust grain.

The interstellar wind enters the Solar system with relative velocity
26.3 km/s and comes from the direction $\lambda_{\text{ecl}}$ $=$
254.7$^{\circ}$ (heliocentric ecliptic longitude) and $\beta_{\text{ecl}}$
$=$ 5.2$^{\circ}$ (heliocentric ecliptic latitude; \citealt{lallement}).
After the passage through various layers caused by magnetohydrodynamic
interaction of the interstellar wind with the solar wind the original
interstellar hydrogen that remains unaffected has the number density
$n_{\text{H~I}}$ $=$ 0.059 g.cm$^{-3}$ \citep{frisch}. The interstellar
helium reaches inner Solar system (neighborhood of the Earth's orbit)
weakly affected by the interaction with the density $n_{\text{He}}$
$=$ 0.015 g.cm$^{-3}$ and the temperature $T_{\text{He}}$ $=$ 6300 K
\citep{frisch}. The temperature of the interstellar helium moving freely
to the inner Solar system is approximately equal to the temperature
of the unaffected interstellar wind. The original interstellar hydrogen
produces the so-called second population by the charge exchange with protons
in the outer heliosheath (between the bowshock and the heliopause).
We used the density $n_{\text{H~II}}$ $=$ 0.059 g.cm$^{-3}$ for
the second population of the interstellar hydrogen after the passage
into the heliosphere \citep{frisch}. The temperatures of two hydrogen
populations are different due to the charge exchange. We used
$T_{\text{H~I}}$ $=$ 6100 K and $T_{\text{H~II}}$ $=$ 16500 K
\citep{frisch}.

When we add the acceleration in Eq. (\ref{ISW}) to Eq. (\ref{speom}),
then we obtain the final equation of motion of the dust grain in
the PCRTBP with the stellar radiation and the interstellar gas flow
\begin{align}\label{mneom}
\frac{d \vec{v}}{dt} = {} & - \frac{\mu}{r^{2}}
      \left ( 1 - \beta \right ) \vec{e}_{\text{R}} -
      \frac{G_{0} M_{\text{P}}}{\vert \vec{r} - \vec{r}_{\text{P}} \vert^{3}}
      (\vec{r} - \vec{r}_{\text{P}}) -
      \frac{G_{0} M_{\text{P}}}{r_{\text{P}}^{3}} \vec{r}_{\text{P}}
\notag \\
& - \beta \frac{\mu}{r^{2}}
      \left ( 1 + \frac{\eta}{\bar{Q}'_{\text{pr}}} \right )
      \left ( \frac{\vec{v} \cdot \vec{e}_{\text{R}}}{c}
      \vec{e}_{\text{R}} + \frac{\vec{v}}{c} \right )
\notag \\
& - \sum_{i = 1}^{N} c_{\text{D}i} \gamma_{i}
      \vert \vec{v} - \vec{v}_{\text{F}} \vert
      \left ( \vec{v} - \vec{v}_{\text{F}} \right ) ~.
\end{align}

\subsection{Secular variations}
\label{subsec:mnsec}

An expansion of the particle's acceleration in Eq. (\ref{ISW})
using Taylor series enables the calculation of secular time derivatives
of the orbital parameters from Gauss's perturbation equations
\citep{dyncd,stab}. The calculated secular time derivatives of orbital
parameters for the stellar radiation and the interstellar gas flow are
\begin{align}\label{mnsecular}
\left ( \frac{da}{dt} \right )_{\text{EF}} = {} & -
      \frac{\beta \mu}{c a \alpha^{3}}
      \left ( 1 + \frac{\eta}{\bar{Q}'_{\text{pr}}} \right )
      \left ( 2 + 3 e^{2} \right )
\notag \\
& - \sum_{i = 1}^{N} \frac{2 c_{0i} \gamma_{i} v_{\text{F}}^{2}
      \sigma_{\text{F}} a^{2} \alpha}{L}
      \left [ 1 + \frac{g_{i} \left ( S^{2} + \alpha I^{2} \right )}
      {v_{\text{F}}^{2} \left ( 1 + \alpha \right )} \right ] ~,
\notag \\
\left ( \frac{de}{dt} \right )_{\text{EF}} = {} & -
      \frac{\beta \mu}{2 c a^{2} \alpha}
      \left ( 1 + \frac{\eta}{\bar{Q}'_{\text{pr}}} \right ) 5 e
\notag \\
& + \sum_{i = 1}^{N} \frac{c_{0i} \gamma_{i} v_{\text{F}} a \alpha}{2 L}
      \left [ 3 I + \frac{\sigma_{\text{F}} g_{i} \alpha^{2}
      \left ( 1 - \alpha \right ) \left ( I^{2} - S^{2} \right )}
      {v_{\text{F}} e \left ( 1 + \alpha \right )} \right ] ~,
\notag \\
\left ( \frac{d \tilde{\omega}}{dt} \right )_{\text{EF}} = {} &
      \sum_{i = 1}^{N} \frac{c_{0i} \gamma_{i} v_{\text{F}} a \alpha S}{2 L}
      \left \{ - \frac{3}{e} +
      \frac{\sigma_{\text{F}} g_{i} I}{v_{\text{F}}}
      \left [ \frac{2 \alpha^{2}}{\left ( 1 + \alpha \right )^{2}} -
      1 \right ] \right \} ~,
\notag \\
\left ( \frac{d \sigma_{\text{b}}}{dt} +
      t \frac{dn}{dt} \right )_{\text{EF}} = {} &
      \sum_{i = 1}^{N} \frac{c_{0i} \gamma_{i} v_{\text{F}} a S}{2 L}
      \Biggl \{ \frac{3 \left ( 1 + e^{2} \right )}{e} -
      \frac{\sigma_{\text{F}} g_{i} \alpha^{2} I}{v_{\text{F}}}
\notag \\
& \times \left [ \frac{2 \alpha^{2}}{\left ( 1 + \alpha \right )^{2}} -
      1 \right ] \Biggr \} ~.
\end{align}
Here
\begin{align}\label{SI}
S &= v_{\text{F}x} \cos \tilde{\omega} + v_{\text{F}y} \sin \tilde{\omega} ~,
\notag \\
I &= - v_{\text{F}x} \sin \tilde{\omega} + v_{\text{F}y} \cos \tilde{\omega} ~.
\end{align}
with $v_{\text{F}x}$ and $v_{\text{F}y}$ denoting the Cartesian components
of the interstellar gas flow velocity vector.
\begin{equation}\label{sigma_F}
\sigma_{\text{F}} = \frac{1}{v_{\text{F}}}
\sqrt{\frac{\mu ( 1 - \beta )}{a ( 1 - e^{2} )}} ~.
\end{equation}
Only the orbits with $\sigma_{\text{F}}^{2}$ negligible in a comparison
with $\sigma_{\text{F}}$ are considered in the expansion. $c_{0i}$ are
the drag coefficients for the dust particle at the rest with
respect to the star. The parameters $g_{i}$ describe dependences
of the drag coefficients on the velocity of dust particle with respect
to the star. For constant drag coefficients hold $g_{i}$ $=$ 1 \citep{dyncd}.
Some of Eqs. (\ref{mnsecular}) are singular in the eccentricity
due to the reasons mentioned after Eqs. (\ref{evolution}).

\begin{figure}[t]
\begin{center}
\includegraphics[width=0.83503521126760563380281690140845\textwidth]{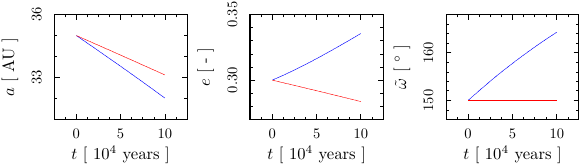}
\end{center}
\caption{A comparison of secular evolutions obtained before (red solid
line) and after (blue solid line) the addition of the interstellar gas flow
to the solar radiation acting on a dust particle with $R$ $=$ 2 $\mu$m,
$\varrho$ $=$ 1 g.cm$^{-3}$, and $\bar{Q}'_{\text{pr}}$ $=$ 1. The evolutions
of the semimajor axis, eccentricity, and longitude of perihelion are
averaged over the orbital period. The secular evolutions are significantly
influenced by the interstellar gas flow at the heliocentric distances
where a capture into the mean motion resonance with the Neptune can
occur.}
\label{fig:cases}
\end{figure}

For the dust particles in the inner Solar system the acceleration from
the interstellar gas flow can be neglected in comparison with the accelerations
from the PR effect and the solar wind. However, in the vicinity
of Neptune's orbit the acceleration from the interstellar gas flow dominates
in the secular evolution of the dust particles. Variations in
the particle's secular evolution caused by the addition of the interstellar
gas flow to the solar radiation are illustrated in Fig. \ref{fig:cases}.
In these plots a dust particle with $R$ $=$ 2 $\mu$m, $\varrho$ $=$
1 g.cm$^{-3}$, and $\bar{Q}'_{\text{pr}}$ $=$ 1 evolves from the initial
conditions $a_{\text{in}}$ $=$ 35 AU, $e_{\text{in}}$ $=$ 0.3,
$\tilde{\omega}$ $=$ 150$^{\circ}$, and $f_{\text{in}}$ $=$ 180$^{\circ}$
without the gravitational influence of the planet. The dependence
of the drag coefficients on the velocity of dust particle with respect
to the star is considered (Eq. \ref{cd}). As can be seen in
Fig. \ref{fig:cases} the influence of the interstellar gas flow cannot
be neglected in the vicinity of Neptune's orbit. On the bound orbits
the addition of the interstellar gas flow causes always faster decrease
of the semimajor axis, the eccentricity can also increase (instead
of the monotonic decrease caused by the solar radiation), and the longitude
of perihelion is not constant (compare Eqs. \ref{spsecular} and
Eqs. \ref{mnsecular}).

Inclination between the Neptune's orbital plane and the interstellar
gas velocity vector is 3.7$^{\circ}$. Therefore, the assumption
that the solved problem is planar is not strictly correct.
The secular time derivative of the inclination caused
by the interstellar gas flow is for orbits with $i$ $\approx$ 0
proportional to $v_{\text{F}z}$ \citep{dyncd} and this velocity component is
small in coordinates with the $xy$ plane lying in the Neptune's orbital plane.
Hence, the inclination can be well approximated by a constant value close
to zero. This is also confirmed by the numerical integration of the equation
of motion. In order to obtain results for the PCRTBP with the PR effect,
solar wind and interstellar gas flow we rotated the interstellar gas velocity
vector into the Neptune's orbital plane around an axis perpendicular to
the interstellar gas velocity vector and lying in the Neptune's orbital plane.

\subsection{Linearization of averaged resonant equations}
\label{subsec:mnlinearization}

The partial derivatives with respect to $a$, $e$, $\tilde{\omega}$, and
$\sigma$ can be calculated using Eqs. (\ref{mnsecular})
(see Eqs. \ref{alphabet} in Appendix \ref{app:coefficients}). The solved
problem does not have the rotational symmetry for the interstellar gas flow and
$\Lambda_{0}$ $\neq$ 0. The $\Lambda_{3}$, $\Lambda_{2}$, $\Lambda_{1}$,
and $\Lambda_{0}$ calculated from Eqs. (\ref{lambda3})-(\ref{lambda0})
determine $\lambda_{i}$ as roots of the quadric equation
Eq. (\ref{quadric}).

\subsection{Numerical checking}
\label{subsec:mncheck}

\begin{figure}[t]
\begin{center}
\includegraphics[width=0.83503521126760563380281690140845\textwidth]{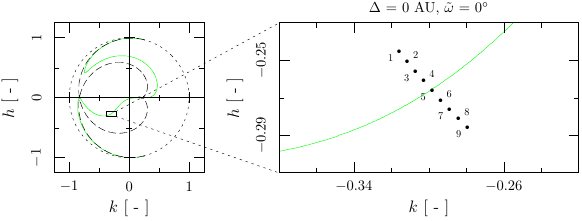}
\end{center}
\caption{The $kh$ plane containing the solutions of resonant
condition in the PCRTBP with solar radiation and interstellar gas flow
at the shift $\Delta$ $=$ 0 and the longitude of perihelion
$\tilde{\omega}$ $=$ 0 for the dust particle with $R$
$=$ 2 $\mu$m, $\varrho$ $=$ 1 g.cm$^{-3}$, and $\bar{Q}'_{\text{pr}}$ $=$ 1
in the exterior mean motion 3/2 resonance with the Neptune.
The initial averaged conditions belonging to the evolutions depicted
in Figs. \ref{fig:mnap} and \ref{fig:mnshift} are shown in the scaled
rectangle region of the $kh$ plane. The same legend as in Fig. \ref{fig:spkh}
is used.}
\label{fig:mnkh}
\end{figure}

\begin{figure}[t]
\begin{center}
\includegraphics[width=0.84700704225352112676056338028169\textwidth]{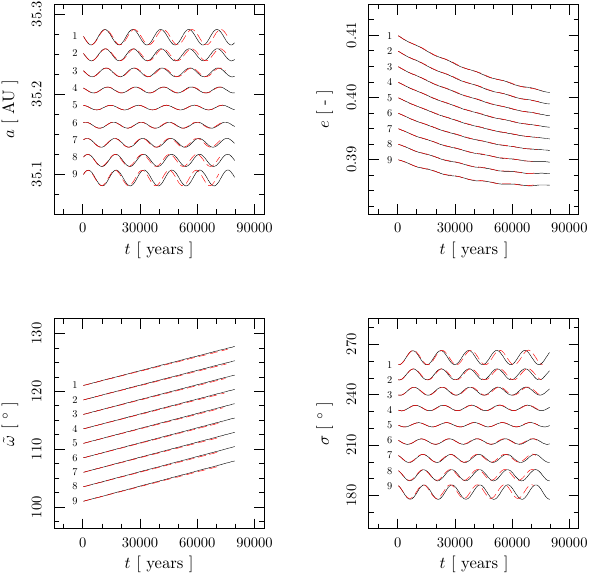}
\end{center}
\caption{The same plots as in Fig. \ref{fig:spap} for the PCRTBP with
different planet, particle, resonance, and non-gravitational effects.
A dust particle with $R$ $=$ 2 $\mu$m, $\varrho$ $=$ 1 g.cm$^{-3}$,
and $\bar{Q}'_{\text{pr}}$ $=$ 1 is initially located in the exterior
mean motion 3/2 resonance with the Neptune under the action of the PR effect,
solar wind and interstellar gas flow. The successive translations
of evolutions $a$, $e$, $\tilde{\omega}$, and $\sigma$ are 2.2 $\times$
10$^{-2}$ AU, 2.5 $\times$ 10$^{-3}$, 3$^{\circ}$, and
10$^{\circ}$, respectively. The evolution 5 is not translated.}
\label{fig:mnap}
\end{figure}

\begin{figure}[t]
\begin{center}
\includegraphics[width=0.90686619718309859154929577464789\textwidth]{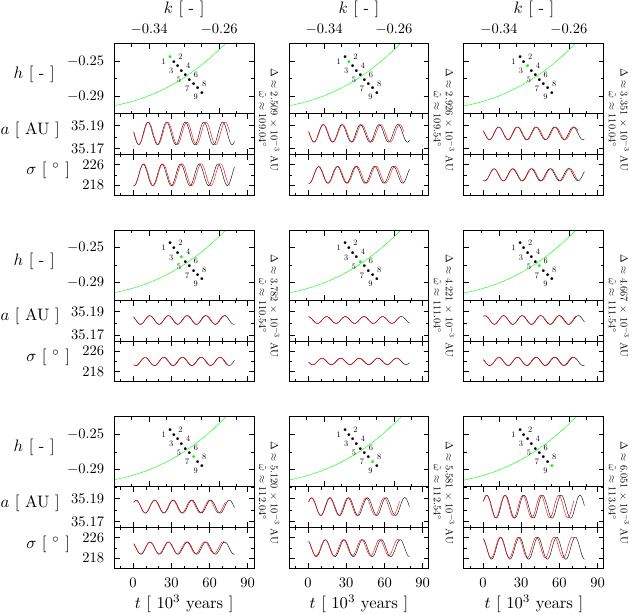}
\end{center}
\caption{The same plots as in Fig. \ref{fig:spshift} depicting data from
the evolutions in Fig. \ref{fig:mnap}. Each plot in addition to the initial
averaged shift contains on the right-hand side also the initial
averaged longitude of perihelion at which was the resonant condition
solved for the evolution marked with the green $kh$ point.}
\label{fig:mnshift}
\end{figure}

The varying longitude of pericenter affects the secular evolution of dust
particles when the interstellar gas is moving through the PCRTBP with
radiation. The phase space containing all evolution for a given mean motion
resonance in the PCRTBP with radiation and interstellar gas flow
has five dimensions ($\beta$, $a$, $e$, $\tilde{\omega}$, $\sigma$).
The resonant condition is in this case
\begin{align}\label{mnaxis}
\frac{da}{dt} = {} & - \frac{2 s a}{L}
\frac{\partial R}{\partial \sigma} -
      \frac{\beta \mu}{c a \alpha^{3}}
      \left ( 1 + \frac{\eta}{\bar{Q}'_{\text{pr}}} \right )
      \left ( 2 + 3 e^{2} \right )
\notag \\
& - \sum_{i = 1}^{N} \frac{2 c_{0i} \gamma_{i} v_{\text{F}}^{2}
      \sigma_{\text{F}} a^{2} \alpha}{L}
      \left [ 1 + \frac{g_{i} \left ( S^{2} + \alpha I^{2} \right )}
      {v_{\text{F}}^{2} \left ( 1 + \alpha \right )} \right ] = 0 ~.
\end{align}
For the sake of simplicity we fixed the semimajor axis in the averaged
phase space by using $\Delta$ $=$ 0 and $\beta$ by choosing one
dust particle with $R$ $=$ 2 $\mu$m, $\varrho$ $=$ 1 g.cm$^{-3}$, and
$\bar{Q}'_{\text{pr}}$ $=$ 1. We solved the resonant condition for
the exact resonance at various longitudes of perihelion in the $kh$ plane.
Interesting property was found. The solution of resonant condition does not
significantly depend on the longitude of perihelion for the considered dust
particle and the interstellar gas in the Solar system. The variations
of the resonant angular variable found from the resonant condition at a given
eccentricity ($e$ $\apprle$ 1 see further) due to the longitude
of perihelion (varying in the interval [0, 2$\pi$]) are typically less than
one degree. The variations in the extreme orientations $S$ $=$
$v_{\text{F}}$ and $I$ $=$ $v_{\text{F}}$ were also compared. This property
was verified for various resonances with the Neptune and holds also
if the Neptune is replaced with the Earth-mass planet. But when
the interstellar gas flow is not considered, then the solution of resonant
condition is different. The solution of the resonant condition with
$\Delta$ $=$ 0 and $\tilde{\omega}$ $=$ 0 for the dust particle with
$R$ $=$ 2 $\mu$m, $\varrho$ $=$ 1 g.cm$^{-3}$, and $\bar{Q}'_{\text{pr}}$
$=$ 1 in the exterior mean motion 3/2 resonance with the Neptune is depicted
in Fig. \ref{fig:mnkh}. The solutions of resonant condition are close
to the collisions for high eccentricities. The high eccentricities are shown
only for a completeness of the depicted solutions, since the approximation
mentioned below Eq. (\ref{sigma_F}) does not hold well for the eccentricities
$\apprge$ 0.8 in the considered problem. The right-hand side panel shows
a region containing the initial averaged conditions for Figs. \ref{fig:mnap}
and \ref{fig:mnshift}.

Fig. \ref{fig:mnap} shows evolutions of the semimajor axis, eccentricity,
longitude of perihelion, and resonant angular variable calculated
numerically from the equation of motion (Eq. \ref{mneom})
and analytically from Eq. (\ref{quadricsolution}) using the initial
averaged conditions in Eqs. (\ref{alphabet}). Initial conditions
for oscular parameters are $\Delta_{\text{in}}$ $=$ 0 AU, $e_{\text{in}}$
$=$ 0.4, $\tilde{\omega}$ $=$ $-$ $\sigma_{\text{in}}$ $q / ( p + q ) $,
and $\sigma_{\text{in}}$ $\in$ $\{ 218^{\circ}, 219^{\circ},
220^{\circ}, ..., 226^{\circ} \}$. The initial true anomalies
of the planet and the particle were zero. The successive translations
of the obtained curves for $a$, $e$, $\tilde{\omega}$, and $\sigma$ are
2.2 $\times$ 10$^{-2}$ AU, 2.5 $\times$ 10$^{-3}$, 3$^{\circ}$, and
10$^{\circ}$, respectively. The zero translation is at the evolution 5.
As in Fig. \ref{fig:spap} the evolutions of $\tilde{\omega}$
without the translation would be shown in the opposite order with
a small separation and the evolutions of the other orbital parameters
would be overlapped. The eccentricity does not approach the universal
eccentricity due to the dependence of the secular time derivatives
of the semimajor axis and the eccentricity on the longitude of perihelion.
The universal eccentricity does not exit for the considered non-gravitational
effects \citep{stab}. Oscillations in the evolution of eccentricity
can be seen. The eccentricity evolves non-monotonically for the evolutions
with the numbers from 7 to 9. During the same number of librations
the longitude of perihelion varies more rapidly in Fig. \ref{fig:mnap}
in comparison with Fig. \ref{fig:spap}.

The linearization frequency well corresponds to the real libration
frequency at the solution of resonant condition as can be seen in
Fig. \ref{fig:mnshift}. The top panel of each plot shows the solution
of resonant condition in the $kh$ plane calculated for the initial
averaged values of the shift and the longitude of perihelion belonging to
the evolution with the green $kh$ point. The librations of the semimajor
axis and the resonant angular variable in Fig. \ref{fig:mnap} are
shown with a different scale in two bottom panels of each plot in
Fig. \ref{fig:mnshift}. The best frequency accordance is obtained at
the evolution 5. The green line of this evolution is between
the $kh$ points 5 and 6.

The parameters giving the linearization solution for the evolution 5
are shown in Appendix \ref{app:lp} (Table \ref{tab:mnlp}). The libration
amplitude increases also for the linearization solution in Table \ref{tab:mnlp}
since the real parts of $\lambda_{3}$ and $\lambda_{4}$ are positive and
the others roots are real. This analytically indicates an instability
for the exterior resonances in the PCRTBP with the solar radiation and
interstellar gas flow. The majority of captures end due to the increase
of libration amplitude \citep{mmrflow}. However, it is possible to find also
the evolutions in the exterior resonances with temporary decreasing libration
amplitude.

For the resonant evolutions with much larger libration amplitudes
of $\sigma$ as those in Figs. \ref{fig:spshift} and \ref{fig:mnshift}
the best frequency accordance is shifted farther from the solution
of resonant condition. In these cases the linearization is not good method
for the calculation of the libration frequency.

\subsection{Periodic solutions}
\label{subsec:mnperiod}

In the PCRTBP with solar radiation and interstellar gas flow such
periodic solutions as those in Sect. \ref{subsec:spperiod} have not been
found. If such periodic solutions exist, then they must have a fixed
(constant) longitude of perihelion in the averaged phase space.
Existence of these periodic orbits is unlikely \citep{stab}. In general,
the condition for the fixed longitude of pericenter depends on the nature
of the non-gravitational effects. As a special case can theoretically
exist the non-gravitational effects without the rotational symmetry that
have periodic solutions with the varying longitude of pericenter. However,
also for such non-gravitational effects the condition for the fixed
longitude of pericenter could give a different set of the periodic orbits.

\section{Conclusion}
\label{sec:conclusion}

We have derived the averaged resonant equations in the PCRTBP with
the non-gravitational effects using summed Lagrange's planetary
equations and Gauss's perturbation equations in few simple steps.
The averaged resonant equations were linearized and solved for
the standard solution in a general form. The planarity of problem
restricts maximal number of evolving parameters describing the orbit
to four. For four evolving parameters the degree of the characteristic
polynomial is four and its analytical solution always exists. This
would not be case for a higher number of evolving parameters.
For problems that include the non-gravitational effects acting with
the rotational symmetry around the star the longitude of pericenter
evolves separately and does not affect remaining three parameters.
The applicability of the linearization solution does not significantly
depend on the variations of the orbit during the averaged
synodic period. This confirms that the evolutions in the mean motion
resonances can be correctly described by the averaged resonant
equations. The actual positions of the planet and the dust
particle in the space are not important for the correct secular
evolution.

The linearization frequency depends most sensitively on the initial
averaged value of the resonant angular variable in the comparison with
dependences on the initial averaged values of the other evolving orbital
parameters. The linearization frequency matches best the real libration
frequency for such initial averaged conditions that are close to the solution
of resonant condition. The solutions of resonant condition are located
in the evolution of the semimajor axis in the minima or maxima.
The minima or maxima of the semimajor axis occur approximately
in the middle of libration in the evolution of the resonant angular
variable.

When the libration amplitude of the resonant angular variable is larger,
then the best frequency accordance is shifted farther from the solutions
of resonant condition. In these cases the validity of linearization
solution disappears. The stationary solutions exist at the solutions
of resonant condition and the resonant angular variable of the stationary
conditions is constant therefore their ``frequency'' obtained from
the linearization solution should be correct.

The linearization solution can be used as an indicator
of the stability of the resonant captures. For example when a small
libration amplitude of the linearization solution always decreases,
then the capture time should be theoretically be infinitely long.
However, when for a stationary solution the linearization solution
gives an increase of the libration amplitude, then this not necessarily
means that the capture time of the stationary solution is theoretically
finite. As an example we can mention the periodic solutions for the exterior
resonances in the PCRTBP with stellar radiation (Sect. \ref{subsec:spperiod}).
In this case a small displacement of the initial conditions from the stationary
solution causes the increase of libration amplitude. The increase of libration
amplitude is proportional to the libration amplitude and this implies that
the zero libration amplitude does not increase. Therefore, it is practically
impossible to choose finite numerical values of the initial conditions for
the linearization solution that should give the constant zero libration
amplitude of the stationary solution. In this case the linearization
solution always gives a small increase of the libration amplitude
regardless of numerical limits of averaging in order to obtain the zero
time derivatives of the orbital parameters.

In the PCRTBP with the PR effect, radial solar wind, and interstellar
gas flow the resonant angular variable satisfying the resonant
condition at a given eccentricity insignificantly depend on
the longitude of perihelion (compare Fig. \ref{fig:mnkh} and
Fig. \ref{fig:mnshift}). For the exterior resonances in this
asymmetrical problem the libration amplitude of the real evolution
usually increases, but the resonant captures with temporary decreasing
libration amplitude exist also.

\appendix

\section{Constant coefficients}
\label{app:coefficients}

This appendix presents coefficients for linearized system of equations
describing the orbital evolutions of the dust particles captured in
the mean motion resonances in the PCRTBP with the PR effect, radial
stellar wind and interstellar gas flow (Eqs. \ref{oscillations}).
\begin{align}\label{alphabet}
A_{\text{c}} = {} & - \frac{s}{L_{0}}
      \frac{\partial R}{\partial \sigma} -
      \frac{2 s a_{0}}{L_{0}}
      \frac{\partial^{2} R}{\partial^{\star} a \partial \sigma} +
      \frac{\beta \mu}{c a_{0}^{2} \alpha_{0}^{3}}
      \left ( 1 + \frac{\eta}{\bar{Q}'_{\text{pr}}} \right )
      \left ( 2 + 3 e_{0}^{2} \right )
\notag \\
& - \sum_{i = 1}^{N} \frac{2 c_{0i} \gamma_{i} v_{\text{F}}^{2}
      \sigma_{\text{F}} a_{0} \alpha_{0}}{L_{0}} \left [ 1 +
      \frac{g_{i} \left ( S_{0}^{2} + \alpha_{0} I_{0}^{2} \right )}
      {v_{\text{F}}^{2} \left ( 1 + \alpha_{0} \right )} \right ] ~,
\notag \\
B_{\text{c}} = {} & - \frac{2 s a_{0}}{L_{0}}
      \frac{\partial^{2} R}{\partial e \partial \sigma} -
      \frac{3 \beta \mu e_{0}}{c a_{0} \alpha_{0}^{5}}
      \left ( 1 + \frac{\eta}{\bar{Q}'_{\text{pr}}} \right )
      \left ( 4 + e_{0}^{2} \right ) -
      \sum_{i = 1}^{N} \frac{2 c_{0i} \gamma_{i}
      \sigma_{\text{F}} a_{0}^{2} e_{0}}{L_{0}}
      \frac{g_{i} \left ( S_{0}^{2} - I_{0}^{2} \right )}
      {\left ( 1 + \alpha_{0} \right )^{2}} ~,
\notag \\
C_{\text{c}} = {} & - \sum_{i = 1}^{N} \frac{4 c_{0i} \gamma_{i}
      \sigma_{\text{F}} a_{0}^{2} \alpha_{0}}{L_{0}}
      \frac{g_{i} S_{0} I_{0} \left ( 1 - \alpha_{0} \right )}
      {1 + \alpha_{0}} ~,
\notag \\
D_{\text{c}} = {} & - \frac{2 s a_{0}}{L_{0}}
      \frac{\partial^{2} R}{\partial \sigma^{2}} ~,
\notag \\
E_{\text{c}} = {} & 0 ~,
\notag \\
F_{\text{c}} = {} & - \frac{2 s a_{0}}{L_{0}}
      \frac{\partial R}{\partial \sigma} -
      \frac{\beta \mu}{c a_{0} \alpha_{0}^{3}}
      \left ( 1 + \frac{\eta}{\bar{Q}'_{\text{pr}}} \right )
      \left ( 2 + 3 e_{0}^{2} \right )
\notag \\
& - \sum_{i = 1}^{N} \frac{2 c_{0i} \gamma_{i} v_{\text{F}}^{2}
      \sigma_{\text{F}} a_{0}^{2} \alpha_{0}}{L_{0}} \left [ 1 +
      \frac{g_{i} \left ( S_{0}^{2} + \alpha_{0} I_{0}^{2} \right )}
      {v_{\text{F}}^{2} \left ( 1 + \alpha_{0} \right )} \right ] ~,
\notag \\
G_{\text{c}} = {} & - \frac{\alpha_{0}}{2 a_{0} L_{0} e_{0}}
      \left [ 1 + s \left ( 1 - \alpha_{0} \right ) \right ]
      \frac{\partial R}{\partial \sigma} +
      \frac{\alpha_{0}}{L_{0} e_{0}}
      \left [ 1 + s \left ( 1 - \alpha_{0} \right ) \right ]
      \frac{\partial^{2} R}{\partial^{\star} a \partial \sigma}
\notag \\
& + \frac{\beta \mu}{c a_{0}^{3} \alpha_{0}}
      \left ( 1 + \frac{\eta}{\bar{Q}'_{\text{pr}}} \right ) 5 e_{0} +
      \sum_{i = 1}^{N} \frac{3 c_{0i} \gamma_{i} v_{\text{F}} \alpha_{0}
      I_{0}}{4 L_{0}} ~,
\notag \\
H_{\text{c}} = {} & - \frac{1}{L_{0} e_{0}^{2} \alpha_{0}}
      \left [ 1 + s \left ( 1 - \alpha_{0} \right ) \right ]
      \frac{\partial R}{\partial \sigma} +
      \frac{s}{L_{0}} \frac{\partial R}{\partial \sigma} +
      \frac{\alpha_{0}}{L_{0} e_{0}}
      \left [ 1 + s \left ( 1 - \alpha_{0} \right ) \right ]
      \frac{\partial^{2} R}{\partial e \partial \sigma}
\notag \\
& - \frac{5 \beta \mu}{2 c a_{0}^{2} \alpha_{0}^{3}}
      \left ( 1 + \frac{\eta}{\bar{Q}'_{\text{pr}}} \right )
\notag \\
& - \sum_{i = 1}^{N} \frac{c_{0i} \gamma_{i} v_{\text{F}}^{2} a_{0}}{2 L_{0}}
      \left \{ \frac{3 e_{0} I_{0}}{v_{\text{F}} \alpha_{0}} +
      \frac{\sigma_{\text{F}} g_{i} \alpha_{0}}{v_{\text{F}}^{2}}
      \left [ 1 - \frac{3}{\left ( 1 + \alpha_{0} \right )^{2}} \right ]
      \left ( S_{0}^{2} - I_{0}^{2} \right ) \right \} ~,
\notag \\
I_{\text{c}} = {} & - \sum_{i = 1}^{N} \frac{c_{0i} \gamma_{i} v_{\text{F}}^{2}
      a_{0} \alpha_{0}}{L_{0}}
      \left [ \frac{3 S_{0}}{2 v_{\text{F}}} +
      \frac{2 \sigma_{\text{F}} g_{i} \alpha_{0}^{2}
      \left ( 1 - \alpha_{0} \right ) S_{0} I_{0}}
      {v_{\text{F}}^{2} e_{0} \left ( 1 + \alpha_{0} \right )} \right ] ~,
\notag \\
J_{\text{c}} = {} & \frac{\alpha_{0}}{L_{0} e_{0}}
      \left [ 1 + s \left ( 1 - \alpha_{0} \right ) \right ]
      \frac{\partial^{2} R}{\partial \sigma^{2}} ~,
\notag \\
K_{\text{c}} = {} & 0 ~,
\notag \\
L_{\text{c}} = {} & \frac{\alpha_{0}}{L_{0} e_{0}}
      \left [ 1 + s \left ( 1 - \alpha_{0} \right ) \right ]
      \frac{\partial R}{\partial \sigma} -
      \frac{5 \beta \mu}{2 c a_{0}^{2} \alpha_{0}}
      \left ( 1 + \frac{\eta}{\bar{Q}'_{\text{pr}}} \right ) e_{0}
\notag \\
& + \sum_{i = 1}^{N} \frac{c_{0i} \gamma_{i} v_{\text{F}}^{2}
      a_{0} \alpha_{0}}{2 L_{0}} \left [ \frac{3 I_{0}}{v_{\text{F}}} -
      \frac{\sigma_{\text{F}} g_{i} \alpha_{0}^{2}
      \left ( 1 - \alpha_{0} \right ) \left ( S_{0}^{2} - I_{0}^{2} \right )}
      {v_{\text{F}}^{2} e_{0} \left ( 1 + \alpha_{0} \right )} \right ] ~,
\notag \\
M_{\text{c}} = {} & - \frac{\alpha_{0}}{2 a_{0} L_{0} e_{0}}
      \frac{\partial R}{\partial e} +
      \frac{\alpha_{0}}{L_{0} e_{0}}
      \frac{\partial^{2} R}{\partial^{\star} a \partial e} -
      \sum_{i = 1}^{N} \frac{3 c_{0i} \gamma_{i} v_{\text{F}}
      \alpha_{0} S_{0}}{4 L_{0} e_{0}} ~,
\notag \\
N_{\text{c}} = {} & - \frac{1}{L_{0} e_{0}^{2} \alpha_{0}}
      \frac{\partial R}{\partial e} +
      \frac{\alpha_{0}}{L_{0} e_{0}}
      \frac{\partial^{2} R}{\partial e^{2}} +
      \sum_{i = 1}^{N} \frac{c_{0i} \gamma_{i} v_{\text{F}}
      a_{0} S_{0}}{L_{0}} \left [ \frac{3}{2 e_{0}^{2} \alpha_{0}} -
      \frac{2 \sigma_{\text{F}} g_{i} \alpha_{0} e_{0} I_{0}}
      {v_{\text{F}} \left ( 1 + \alpha_{0} \right )^{3}} \right ] ~,
\notag \\
O_{\text{c}} = {} & - \sum_{i = 1}^{N} \frac{c_{0i} \gamma_{i} v_{\text{F}}
      a_{0} \alpha_{0}}{2 L_{0}} \left \{ \frac{3 I_{0}}{e_{0}} +
      \frac{\sigma_{\text{F}} g_{i}}{v_{\text{F}}}
      \left [ \frac{2 \alpha_{0}^{2}}{\left ( 1 + \alpha_{0} \right )^{2}} -
      1 \right ] \left ( S_{0}^{2} - I_{0}^{2} \right ) \right \} ~,
\notag \\
P_{\text{c}} = {} & \frac{\alpha_{0}}{L_{0} e_{0}}
      \frac{\partial^{2} R}{\partial \sigma \partial e} ~,
\notag \\
Q_{\text{c}} = {} & 0 ~,
\notag \\
R_{\text{c}} = {} & \frac{\alpha_{0}}{L_{0} e_{0}}
      \frac{\partial R}{\partial e} +
      \sum_{i = 1}^{N} \frac{c_{0i} \gamma_{i} v_{\text{F}}
      a_{0} \alpha_{0} S_{0}}{2 L_{0}} \left \{ - \frac{3}{e_{0}} +
      \frac{\sigma_{\text{F}} g_{i} I_{0}}{v_{\text{F}}}
      \left [ \frac{2 \alpha_{0}^{2}}{\left ( 1 + \alpha_{0} \right )^{2}} -
      1 \right ] \right \} ~,
\notag \\
S_{\text{c}} = {} & \frac{\alpha_{0}}{2 a_{0} L_{0} e_{0}}
      \left [ 1 + s \left ( 1 - \alpha_{0} \right ) \right ]
      \frac{\partial R}{\partial e} -
      \frac{\alpha_{0}}{L_{0} e_{0}}
      \left [ 1 + s \left ( 1 - \alpha_{0} \right ) \right ]
      \frac{\partial^{2} R}{\partial^{\star} a \partial e} +
      \frac{s}{L_{0}} \frac{\partial R}{\partial^{\star} a} +
      \frac{2 s a_{0}}{L_{0}}
      \frac{\partial^{2} R}{\partial^{\star} a^{2}}
\notag \\
& + \frac{3 s n_{0}}{2 a_{0}} - s \sum_{i = 1}^{N} \frac{3 c_{0i}
      \gamma_{i} v_{\text{F}} e_{0} S_{0}}{2 L_{0}} +
      \left [ 1 + s \left ( 1 - \alpha_{0} \right ) \right ]
      \sum_{i = 1}^{N} \frac{3 c_{0i} \gamma_{i} v_{\text{F}}
      \alpha_{0} S_{0}}{4 L_{0} e_{0}} ~,
\notag \\
T_{\text{c}} = {} & \frac{1}{L_{0} e_{0}^{2} \alpha_{0}}
      \left [ 1 + s \left ( 1 - \alpha_{0} \right ) \right ]
      \frac{\partial R}{\partial e} - \frac{\alpha_{0}}{L_{0} e_{0}}
      \left [ 1 + s \left ( 1 - \alpha_{0} \right ) \right ]
      \frac{\partial^{2} R}{\partial e^{2}} -
      \frac{s}{L_{0}} \frac{\partial R}{\partial e} +
      \frac{2 s a_{0}}{L_{0}}
      \frac{\partial^{2} R}{\partial e \partial^{\star} a}
\notag \\
& - s \sum_{i = 1}^{N} \frac{c_{0i} \gamma_{i} v_{\text{F}}
      a_{0} S_{0}}{2 L_{0}} \left \{ 3 +
      \frac{\sigma_{\text{F}} g_{i} e_{0} I_{0}}{v_{\text{F}}}
      \left [ \frac{2 \alpha_{0}^{2}}{\left ( 1 + \alpha_{0} \right )^{2}} -
      1 \right ] \right \}
\notag \\
& + \left [ 1 + s \left ( 1 - \alpha_{0} \right ) \right ]
      \sum_{i = 1}^{N} \frac{c_{0i} \gamma_{i} v_{\text{F}}
      a_{0} S_{0}}{L_{0}} \left [ - \frac{3}{2 e_{0}^{2} \alpha_{0}} +
      \frac{2 \sigma_{\text{F}} g_{i} e_{0} \alpha_{0} I_{0}}
      {v_{\text{F}} \left ( 1 + \alpha_{0} \right )^{3}} \right ] ~,
\notag \\
U_{\text{c}} = {} & - s \sum_{i = 1}^{N} \frac{3 c_{0i} \gamma_{i} v_{\text{F}}
      a_{0} e_{0} I_{0}}{L_{0}}
\notag \\
& + \left [ 1 + s \left ( 1 - \alpha_{0} \right ) \right ]
      \sum_{i = 1}^{N} \frac{c_{0i} \gamma_{i} v_{\text{F}}^{2}
      a_{0} \alpha_{0}}{2 L_{0}} \left \{ \frac{3 I_{0}}{v_{\text{F}} e_{0}} +
      \frac{\sigma_{\text{F}} g_{i}}{v_{\text{F}}^{2}}
      \left [ \frac{2 \alpha_{0}^{2}}{\left ( 1 + \alpha_{0} \right )^{2}} -
      1 \right ] \left ( S_{0}^{2} - I_{0}^{2} \right ) \right \} ~,
\notag \\
V_{\text{c}} = {} & - \frac{\alpha_{0}}{L_{0} e_{0}}
      \left [ 1 + s \left ( 1 - \alpha_{0} \right ) \right ]
      \frac{\partial^{2} R}{\partial \sigma \partial e} +
      \frac{2 s a_{0}}{L_{0}}
      \frac{\partial^{2} R}{\partial \sigma \partial^{\star} a} ~,
\notag \\
W_{\text{c}} = {} & 0 ~,
\notag \\
X_{\text{c}} = {} & - \frac{\alpha_{0}}{L_{0} e_{0}}
      \left [ 1 + s \left ( 1 - \alpha_{0} \right ) \right ]
      \frac{\partial R}{\partial e} +
      \frac{2 s a_{0}}{L_{0}}
      \frac{\partial R}{\partial^{\star} a} +
      \frac{p + q}{q} n_{\text{P}} - s n_{0} -
      s \sum_{i = 1}^{N} \frac{3 c_{0i} \gamma_{i} v_{\text{F}}
      a_{0} e_{0} S_{0}}{L_{0}}
\notag \\
& + \left [ 1 + s \left ( 1 - \alpha_{0} \right ) \right ]
      \sum_{i = 1}^{N} \frac{c_{0i} \gamma_{i} v_{\text{F}}
      a_{0} \alpha_{0} S_{0}}{2 L_{0}} \left \{ \frac{3}{e_{0}} -
      \frac{\sigma_{\text{F}} g_{i} I_{0}}{v_{\text{F}}}
      \left [ \frac{2 \alpha_{0}^{2}}{\left ( 1 + \alpha_{0} \right )^{2}} -
      1 \right ] \right \} ~.
\end{align}

\section{Constants in the separated equations}
\label{app:separation}

One possible way how we can obtain the separated equation for $\delta_{a}$
in Eqs. (\ref{complete}) is to calculate the following time derivatives
of the first equation in Eqs. (\ref{oscillations}).
\begin{alignat}{14}\label{sepnot}
{} & \dot{\delta}_{a} {} & {} & = {} &
      {} & A_{\text{c}} \delta_{a} {} & {} & + {} &
      {} & B_{\text{c}} \delta_{e} {} & {} & + {} &
      {} & C_{\text{c}} \delta_{\tilde{\omega}} {} & {} & + {} &
      {} & D_{\text{c}} \delta_{\sigma} {} & {} & + {} &
      {} & E_{\text{c}} t {} & {} & + {} & {} & F {} & ~, {} &
\notag \\
{} & \ddot{\delta}_{a} {} & {} & = {} &
      {} & \alpha_{2} \delta_{a} {} & {} & + {} &
      {} & \beta_{2} \delta_{e} {} & {} & + {} &
      {} & \gamma_{2} \delta_{\tilde{\omega}} {} & {} & + {} &
      {} & \delta_{2} \delta_{\sigma} {} & {} & + {} &
      {} & \epsilon_{2} t {} & {} & + {} & {} & \zeta_{2} {} & ~, {} &
\notag \\
{} & \dddot{\delta}_{a} {} & {} & = {} &
      {} & \alpha_{3} \delta_{a} {} & {} & + {} &
      {} & \beta_{3} \delta_{e} {} & {} & + {} &
      {} & \gamma_{3} \delta_{\tilde{\omega}} {} & {} & + {} &
      {} & \delta_{3} \delta_{\sigma} {} & {} & + {} &
      {} & \epsilon_{3} t {} & {} & + {} & {} & \zeta_{3} {} & ~, {} &
\notag \\
{} & \ddddot{\delta}_{a} {} & {} & = {} &
      {} & \alpha_{4} \delta_{a} {} & {} & + {} &
      {} & \beta_{4} \delta_{e} {} & {} & + {} &
      {} & \gamma_{4} \delta_{\tilde{\omega}} {} & {} & + {} &
      {} & \delta_{4} \delta_{\sigma} {} & {} & + {} &
      {} & \epsilon_{4} t {} & {} & + {} & {} & \zeta_{4} {} & ~, {} &
\end{alignat}
here $\alpha_{l}$, $\beta_{l}$, $\gamma_{l}$, $\delta_{l}$, $\epsilon_{l}$,
and $\zeta_{l}$ for $l$ $=$ 2, 3, 4 are determined by the constants
in Eqs. (\ref{oscillations}) as follows
\begin{alignat}{5}\label{sep2}
\alpha_2 {} & = {} & {} & \left ( \begin{array}{cccc}
      A_{\text{c}} & B_{\text{c}} & C_{\text{c}} & D_{\text{c}}
      \end{array} \right )
      \left ( \begin{array}{c}
      A_{\text{c}} \\
      G_{\text{c}} \\
      M_{\text{c}} \\
      S_{\text{c}} \\
      \end{array} \right ) {} & ~, ~~
\beta_2 {} & = {} & {} & \left ( \begin{array}{cccc}
      A_{\text{c}} & B_{\text{c}} & C_{\text{c}} & D_{\text{c}}
      \end{array} \right )
      \left ( \begin{array}{c}
      B_{\text{c}} \\
      H_{\text{c}} \\
      N_{\text{c}} \\
      T_{\text{c}} \\
      \end{array} \right ) ~, {} & {} &
\notag \\
\gamma_2 {} & = {} & {} & \left ( \begin{array}{cccc}
      A_{\text{c}} & B_{\text{c}} & C_{\text{c}} & D_{\text{c}}
      \end{array} \right )
      \left ( \begin{array}{c}
      C_{\text{c}} \\
      I_{\text{c}} \\
      O_{\text{c}} \\
      U_{\text{c}} \\
      \end{array} \right ) {} & ~, ~~
\delta_2 {} & = {} & {} & \left ( \begin{array}{cccc}
      A_{\text{c}} & B_{\text{c}} & C_{\text{c}} & D_{\text{c}}
      \end{array} \right )
      \left ( \begin{array}{c}
      D_{\text{c}} \\
      J_{\text{c}} \\
      P_{\text{c}} \\
      V_{\text{c}} \\
      \end{array} \right ) ~, {} & {} &
\notag \\
\epsilon_2 {} & = {} & {} & \left ( \begin{array}{cccc}
      A_{\text{c}} & B_{\text{c}} & C_{\text{c}} & D_{\text{c}}
      \end{array} \right )
      \left ( \begin{array}{c}
      E_{\text{c}} \\
      K_{\text{c}} \\
      Q_{\text{c}} \\
      W_{\text{c}} \\
      \end{array} \right ) {} & ~, ~~
\zeta_2 {} & = {} & {} & \left ( \begin{array}{cccc}
      A_{\text{c}} & B_{\text{c}} & C_{\text{c}} & D_{\text{c}}
      \end{array} \right )
      \left ( \begin{array}{c}
      F_{\text{c}} \\
      L_{\text{c}} \\
      R_{\text{c}} \\
      X_{\text{c}} \\
      \end{array} \right ) + E_{\text{c}} ~. {} & {} &
\end{alignat}
\begin{alignat}{1}\label{sep3}
\alpha_3 = {} & \left ( \begin{array}{cccc}
      A_{\text{c}} & B_{\text{c}} & C_{\text{c}} & D_{\text{c}}
      \end{array} \right )
      \left ( \begin{array}{cccc}
      A_{\text{c}} & B_{\text{c}} & C_{\text{c}} & D_{\text{c}} \\
      G_{\text{c}} & H_{\text{c}} & I_{\text{c}} & J_{\text{c}} \\
      M_{\text{c}} & N_{\text{c}} & O_{\text{c}} & P_{\text{c}} \\
      S_{\text{c}} & T_{\text{c}} & U_{\text{c}} & V_{\text{c}} \\
      \end{array} \right )
      \left ( \begin{array}{c}
      A_{\text{c}} \\
      G_{\text{c}} \\
      M_{\text{c}} \\
      S_{\text{c}} \\
      \end{array} \right ) ~,
\notag \\
\beta_3 = {} & \left ( \begin{array}{cccc}
      A_{\text{c}} & B_{\text{c}} & C_{\text{c}} & D_{\text{c}}
      \end{array} \right )
      \left ( \begin{array}{cccc}
      A_{\text{c}} & B_{\text{c}} & C_{\text{c}} & D_{\text{c}} \\
      G_{\text{c}} & H_{\text{c}} & I_{\text{c}} & J_{\text{c}} \\
      M_{\text{c}} & N_{\text{c}} & O_{\text{c}} & P_{\text{c}} \\
      S_{\text{c}} & T_{\text{c}} & U_{\text{c}} & V_{\text{c}} \\
      \end{array} \right )
      \left ( \begin{array}{c}
      B_{\text{c}} \\
      H_{\text{c}} \\
      N_{\text{c}} \\
      T_{\text{c}} \\
      \end{array} \right ) ~,
\notag \\
\gamma_3 = {} & \left ( \begin{array}{cccc}
      A_{\text{c}} & B_{\text{c}} & C_{\text{c}} & D_{\text{c}}
      \end{array} \right )
      \left ( \begin{array}{cccc}
      A_{\text{c}} & B_{\text{c}} & C_{\text{c}} & D_{\text{c}} \\
      G_{\text{c}} & H_{\text{c}} & I_{\text{c}} & J_{\text{c}} \\
      M_{\text{c}} & N_{\text{c}} & O_{\text{c}} & P_{\text{c}} \\
      S_{\text{c}} & T_{\text{c}} & U_{\text{c}} & V_{\text{c}} \\
      \end{array} \right )
      \left ( \begin{array}{c}
      C_{\text{c}} \\
      I_{\text{c}} \\
      O_{\text{c}} \\
      U_{\text{c}} \\
      \end{array} \right ) ~,
\notag \\
\delta_3 = {} & \left ( \begin{array}{cccc}
      A_{\text{c}} & B_{\text{c}} & C_{\text{c}} & D_{\text{c}}
      \end{array} \right )
      \left ( \begin{array}{cccc}
      A_{\text{c}} & B_{\text{c}} & C_{\text{c}} & D_{\text{c}} \\
      G_{\text{c}} & H_{\text{c}} & I_{\text{c}} & J_{\text{c}} \\
      M_{\text{c}} & N_{\text{c}} & O_{\text{c}} & P_{\text{c}} \\
      S_{\text{c}} & T_{\text{c}} & U_{\text{c}} & V_{\text{c}} \\
      \end{array} \right )
      \left ( \begin{array}{c}
      D_{\text{c}} \\
      J_{\text{c}} \\
      P_{\text{c}} \\
      V_{\text{c}} \\
      \end{array} \right ) ~,
\notag \\
\epsilon_3 = {} & \left ( \begin{array}{cccc}
      A_{\text{c}} & B_{\text{c}} & C_{\text{c}} & D_{\text{c}}
      \end{array} \right )
      \left ( \begin{array}{cccc}
      A_{\text{c}} & B_{\text{c}} & C_{\text{c}} & D_{\text{c}} \\
      G_{\text{c}} & H_{\text{c}} & I_{\text{c}} & J_{\text{c}} \\
      M_{\text{c}} & N_{\text{c}} & O_{\text{c}} & P_{\text{c}} \\
      S_{\text{c}} & T_{\text{c}} & U_{\text{c}} & V_{\text{c}} \\
      \end{array} \right )
      \left ( \begin{array}{c}
      E_{\text{c}} \\
      K_{\text{c}} \\
      Q_{\text{c}} \\
      W_{\text{c}} \\
      \end{array} \right ) ~,
\notag \\
\zeta_3 = {} & \left ( \begin{array}{cccc}
      A_{\text{c}} & B_{\text{c}} & C_{\text{c}} & D_{\text{c}}
      \end{array} \right )
      \left ( \begin{array}{cccc}
      A_{\text{c}} & B_{\text{c}} & C_{\text{c}} & D_{\text{c}} \\
      G_{\text{c}} & H_{\text{c}} & I_{\text{c}} & J_{\text{c}} \\
      M_{\text{c}} & N_{\text{c}} & O_{\text{c}} & P_{\text{c}} \\
      S_{\text{c}} & T_{\text{c}} & U_{\text{c}} & V_{\text{c}} \\
      \end{array} \right )
      \left ( \begin{array}{c}
      F_{\text{c}} \\
      L_{\text{c}} \\
      R_{\text{c}} \\
      X_{\text{c}} \\
      \end{array} \right ) + \left ( \begin{array}{cccc}
      A_{\text{c}} & B_{\text{c}} & C_{\text{c}} & D_{\text{c}}
      \end{array} \right )
      \left ( \begin{array}{c}
      E_{\text{c}} \\
      K_{\text{c}} \\
      Q_{\text{c}} \\
      W_{\text{c}} \\
      \end{array} \right ) ~.
\end{alignat}
\begin{alignat}{1}\label{sep4}
\alpha_4 = {} & \left ( \begin{array}{cccc}
      A_{\text{c}} & B_{\text{c}} & C_{\text{c}} & D_{\text{c}}
      \end{array} \right )
      \left ( \begin{array}{cccc}
      A_{\text{c}} & B_{\text{c}} & C_{\text{c}} & D_{\text{c}} \\
      G_{\text{c}} & H_{\text{c}} & I_{\text{c}} & J_{\text{c}} \\
      M_{\text{c}} & N_{\text{c}} & O_{\text{c}} & P_{\text{c}} \\
      S_{\text{c}} & T_{\text{c}} & U_{\text{c}} & V_{\text{c}} \\
      \end{array} \right )
      \left ( \begin{array}{cccc}
      A_{\text{c}} & B_{\text{c}} & C_{\text{c}} & D_{\text{c}} \\
      G_{\text{c}} & H_{\text{c}} & I_{\text{c}} & J_{\text{c}} \\
      M_{\text{c}} & N_{\text{c}} & O_{\text{c}} & P_{\text{c}} \\
      S_{\text{c}} & T_{\text{c}} & U_{\text{c}} & V_{\text{c}} \\
      \end{array} \right )
      \left ( \begin{array}{c}
      A_{\text{c}} \\
      G_{\text{c}} \\
      M_{\text{c}} \\
      S_{\text{c}} \\
      \end{array} \right ) ~,
\notag \\
\beta_4 = {} & \left ( \begin{array}{cccc}
      A_{\text{c}} & B_{\text{c}} & C_{\text{c}} & D_{\text{c}}
      \end{array} \right )
      \left ( \begin{array}{cccc}
      A_{\text{c}} & B_{\text{c}} & C_{\text{c}} & D_{\text{c}} \\
      G_{\text{c}} & H_{\text{c}} & I_{\text{c}} & J_{\text{c}} \\
      M_{\text{c}} & N_{\text{c}} & O_{\text{c}} & P_{\text{c}} \\
      S_{\text{c}} & T_{\text{c}} & U_{\text{c}} & V_{\text{c}} \\
      \end{array} \right )
      \left ( \begin{array}{cccc}
      A_{\text{c}} & B_{\text{c}} & C_{\text{c}} & D_{\text{c}} \\
      G_{\text{c}} & H_{\text{c}} & I_{\text{c}} & J_{\text{c}} \\
      M_{\text{c}} & N_{\text{c}} & O_{\text{c}} & P_{\text{c}} \\
      S_{\text{c}} & T_{\text{c}} & U_{\text{c}} & V_{\text{c}} \\
      \end{array} \right )
      \left ( \begin{array}{c}
      B_{\text{c}} \\
      H_{\text{c}} \\
      N_{\text{c}} \\
      T_{\text{c}} \\
      \end{array} \right ) ~,
\notag \\
\gamma_4 = {} & \left ( \begin{array}{cccc}
      A_{\text{c}} & B_{\text{c}} & C_{\text{c}} & D_{\text{c}}
      \end{array} \right )
      \left ( \begin{array}{cccc}
      A_{\text{c}} & B_{\text{c}} & C_{\text{c}} & D_{\text{c}} \\
      G_{\text{c}} & H_{\text{c}} & I_{\text{c}} & J_{\text{c}} \\
      M_{\text{c}} & N_{\text{c}} & O_{\text{c}} & P_{\text{c}} \\
      S_{\text{c}} & T_{\text{c}} & U_{\text{c}} & V_{\text{c}} \\
      \end{array} \right )
      \left ( \begin{array}{cccc}
      A_{\text{c}} & B_{\text{c}} & C_{\text{c}} & D_{\text{c}} \\
      G_{\text{c}} & H_{\text{c}} & I_{\text{c}} & J_{\text{c}} \\
      M_{\text{c}} & N_{\text{c}} & O_{\text{c}} & P_{\text{c}} \\
      S_{\text{c}} & T_{\text{c}} & U_{\text{c}} & V_{\text{c}} \\
      \end{array} \right )
      \left ( \begin{array}{c}
      C_{\text{c}} \\
      I_{\text{c}} \\
      O_{\text{c}} \\
      U_{\text{c}} \\
      \end{array} \right ) ~,
\notag \\
\delta_4 = {} & \left ( \begin{array}{cccc}
      A_{\text{c}} & B_{\text{c}} & C_{\text{c}} & D_{\text{c}}
      \end{array} \right )
      \left ( \begin{array}{cccc}
      A_{\text{c}} & B_{\text{c}} & C_{\text{c}} & D_{\text{c}} \\
      G_{\text{c}} & H_{\text{c}} & I_{\text{c}} & J_{\text{c}} \\
      M_{\text{c}} & N_{\text{c}} & O_{\text{c}} & P_{\text{c}} \\
      S_{\text{c}} & T_{\text{c}} & U_{\text{c}} & V_{\text{c}} \\
      \end{array} \right )
      \left ( \begin{array}{cccc}
      A_{\text{c}} & B_{\text{c}} & C_{\text{c}} & D_{\text{c}} \\
      G_{\text{c}} & H_{\text{c}} & I_{\text{c}} & J_{\text{c}} \\
      M_{\text{c}} & N_{\text{c}} & O_{\text{c}} & P_{\text{c}} \\
      S_{\text{c}} & T_{\text{c}} & U_{\text{c}} & V_{\text{c}} \\
      \end{array} \right )
      \left ( \begin{array}{c}
      D_{\text{c}} \\
      J_{\text{c}} \\
      P_{\text{c}} \\
      V_{\text{c}} \\
      \end{array} \right ) ~,
\notag \\
\epsilon_4 = {} & \left ( \begin{array}{cccc}
      A_{\text{c}} & B_{\text{c}} & C_{\text{c}} & D_{\text{c}}
      \end{array} \right )
      \left ( \begin{array}{cccc}
      A_{\text{c}} & B_{\text{c}} & C_{\text{c}} & D_{\text{c}} \\
      G_{\text{c}} & H_{\text{c}} & I_{\text{c}} & J_{\text{c}} \\
      M_{\text{c}} & N_{\text{c}} & O_{\text{c}} & P_{\text{c}} \\
      S_{\text{c}} & T_{\text{c}} & U_{\text{c}} & V_{\text{c}} \\
      \end{array} \right )
      \left ( \begin{array}{cccc}
      A_{\text{c}} & B_{\text{c}} & C_{\text{c}} & D_{\text{c}} \\
      G_{\text{c}} & H_{\text{c}} & I_{\text{c}} & J_{\text{c}} \\
      M_{\text{c}} & N_{\text{c}} & O_{\text{c}} & P_{\text{c}} \\
      S_{\text{c}} & T_{\text{c}} & U_{\text{c}} & V_{\text{c}} \\
      \end{array} \right )
      \left ( \begin{array}{c}
      E_{\text{c}} \\
      K_{\text{c}} \\
      Q_{\text{c}} \\
      W_{\text{c}} \\
      \end{array} \right ) ~,
\notag \\
\zeta_4 = {} & \left ( \begin{array}{cccc}
      A_{\text{c}} & B_{\text{c}} & C_{\text{c}} & D_{\text{c}}
      \end{array} \right )
      \left ( \begin{array}{cccc}
      A_{\text{c}} & B_{\text{c}} & C_{\text{c}} & D_{\text{c}} \\
      G_{\text{c}} & H_{\text{c}} & I_{\text{c}} & J_{\text{c}} \\
      M_{\text{c}} & N_{\text{c}} & O_{\text{c}} & P_{\text{c}} \\
      S_{\text{c}} & T_{\text{c}} & U_{\text{c}} & V_{\text{c}} \\
      \end{array} \right )
      \left ( \begin{array}{cccc}
      A_{\text{c}} & B_{\text{c}} & C_{\text{c}} & D_{\text{c}} \\
      G_{\text{c}} & H_{\text{c}} & I_{\text{c}} & J_{\text{c}} \\
      M_{\text{c}} & N_{\text{c}} & O_{\text{c}} & P_{\text{c}} \\
      S_{\text{c}} & T_{\text{c}} & U_{\text{c}} & V_{\text{c}} \\
      \end{array} \right )
      \left ( \begin{array}{c}
      F_{\text{c}} \\
      L_{\text{c}} \\
      R_{\text{c}} \\
      X_{\text{c}} \\
      \end{array} \right )
\notag \\
{} & + \left ( \begin{array}{cccc}
      A_{\text{c}} & B_{\text{c}} & C_{\text{c}} & D_{\text{c}}
      \end{array} \right )
      \left ( \begin{array}{cccc}
      A_{\text{c}} & B_{\text{c}} & C_{\text{c}} & D_{\text{c}} \\
      G_{\text{c}} & H_{\text{c}} & I_{\text{c}} & J_{\text{c}} \\
      M_{\text{c}} & N_{\text{c}} & O_{\text{c}} & P_{\text{c}} \\
      S_{\text{c}} & T_{\text{c}} & U_{\text{c}} & V_{\text{c}} \\
      \end{array} \right )
      \left ( \begin{array}{c}
      E_{\text{c}} \\
      K_{\text{c}} \\
      Q_{\text{c}} \\
      W_{\text{c}} \\
      \end{array} \right ) ~.
\end{alignat}
We can substitute Eqs. (\ref{sepnot}) in the first equation in
Eqs. (\ref{complete}). Now, when we realize that the first equation
in Eqs. (\ref{complete}) should be valid for arbitrary variations, then
we obtain the following system of equations
\begin{alignat}{11}\label{sepsys}
{} & \alpha_{4} {} &
      {} & + {} &
      {} & \Lambda_{a 3} ~\alpha_{3} {} &
      {} & + {} &
      {} & \Lambda_{a 2} ~\alpha_{2} {} &
      {} & + {} &
      {} & \Lambda_{a 1} ~A_{\text{c}} {} &
      {} & + {} &
      {} & \Lambda_{a 0} {} & {} & = {} & {} & 0 ~,
\notag \\
{} & \beta_{4} {} &
      {} & + {} &
      {} & \Lambda_{a 3} ~\beta_{3} {} &
      {} & + {} &
      {} & \Lambda_{a 2} ~\beta_{2} {} &
      {} & + {} &
      {} & \Lambda_{a 1} ~B_{\text{c}} {} & {} & = {} & {} & 0 ~,
\notag \\
{} & \gamma_{4} {} &
      {} & + {} &
      {} & \Lambda_{a 3} ~\gamma_{3} {} &
      {} & + {} &
      {} & \Lambda_{a 2} ~\gamma_{2} {} &
      {} & + {} &
      {} & \Lambda_{a 1} ~C_{\text{c}} {} & {} & = {} & {} & 0 ~,
\notag \\
{} & \delta_{4} {} &
      {} & + {} &
      {} & \Lambda_{a 3} ~\delta_{3} {} &
      {} & + {} &
      {} & \Lambda_{a 2} ~\delta_{2} {} &
      {} & + {} &
      {} & \Lambda_{a 1} ~D_{\text{c}} {} & {} & = {} & {} & 0 ~,
\notag \\
{} & \epsilon_{4} {} &
      {} & + {} &
      {} & \Lambda_{a 3} ~\epsilon_{3} {} &
      {} & + {} &
      {} & \Lambda_{a 2} ~\epsilon_{2} {} &
      {} & + {} &
      {} & \Lambda_{a 1} ~E_{\text{c}} {} &
      {} & + {} &
      {} & \Lambda_{a t} {} & {} & = {} & {} & 0 ~,
\notag \\
{} & \zeta_{4} {} &
      {} & + {} &
      {} & \Lambda_{a 3} ~\zeta_{3} {} &
      {} & + {} &
      {} & \Lambda_{a 2} ~\zeta_{2} {} &
      {} & + {} &
      {} & \Lambda_{a 1} ~F_{\text{c}} {} &
      {} & + {} &
      {} & \Lambda_{a} {} & {} & = {} & {} & 0 ~.
\end{alignat}
The solution of Eqs. (\ref{sepsys}) gives unknown constants in the separated
equation for $\delta_{a}$.

\section{Equivalency in the symmetrical case when the evolution of longitude
of pericenter is not considered}
\label{app:equivalency}

The solutions in Eq. (\ref{cubicsolution}) for $\delta_{a}$,
$\delta_{e}$, and $\delta_{\sigma}$ are equivalent with the solutions
of the following system of equations
\begin{alignat}{12}\label{equivalent}
{} & \dot{\delta}_{a} {} & {} & = {} &
      {} & A_{\text{c}} \delta_{a} {} & {} & + {} &
      {} & B_{\text{c}} \delta_{e} {} & {} & + {} &
      {} & D_{\text{c}} \delta_{\sigma} {} & {} & + {} &
      {} & E_{\text{c}} t {} & {} & + {} & {} & F {} & ~, {} &
\notag \\
{} & \dot{\delta}_{e} {} & {} & = {} &
      {} & G_{\text{c}} \delta_{a} {} & {} & + {} &
      {} & H_{\text{c}} \delta_{e} {} & {} & + {} &
      {} & J_{\text{c}} \delta_{\sigma} {} & {} & + {} &
      {} & K_{\text{c}} t {} & {} & + {} & {} & L {} & ~, {} &
\notag \\
{} & \dot{\delta}_{\sigma} {} & {} & = {} &
      {} & S_{\text{c}} \delta_{a} {} & {} & + {} &
      {} & T_{\text{c}} \delta_{e} {} & {} & + {} &
      {} & V_{\text{c}} \delta_{\sigma} {} & {} & + {} &
      {} & W_{\text{c}} t {} & {} & + {} & {} & X {} & ~. {} &
\end{alignat}
The separated equations of this system are
\begin{alignat}{13}\label{notation}
{} & \dddot{\delta}_{a} {} &
      {} & + {} &
      {} & \Lambda_{a 3} ~\ddot{\delta}_{a} {} &
      {} & + {} &
      {} & \Lambda_{a 2} ~\dot{\delta}_{a} {} &
      {} & + {} &
      {} & \Lambda_{a 1} ~\delta_{a} {} &
      {} & + {} &
      {} & \Lambda_{a t}^{\star} ~t {} &
      {} & + {} &
      {} & \Lambda_{a}^{\star} {} & {} & = {} & {} & 0 ~,
\notag \\
{} & \dddot{\delta}_{e} {} &
      {} & + {} &
      {} & \Lambda_{e 3} ~\ddot{\delta}_{e} {} &
      {} & + {} &
      {} & \Lambda_{e 2} ~\dot{\delta}_{e} {} &
      {} & + {} &
      {} & \Lambda_{e 1} ~\delta_{e} {} &
      {} & + {} &
      {} & \Lambda_{e t}^{\star} ~t {} &
      {} & + {} &
      {} & \Lambda_{e}^{\star} {} & {} & = {} & {} & 0 ~,
\notag \\
{} & \dddot{\delta}_{\sigma} {} &
      {} & + {} &
      {} & \Lambda_{\sigma 3} ~\ddot{\delta}_{\sigma} {} &
      {} & + {} &
      {} & \Lambda_{\sigma 2} ~\dot{\delta}_{\sigma} {} &
      {} & + {} &
      {} & \Lambda_{\sigma 1} ~\delta_{\sigma} {} &
      {} & + {} &
      {} & \Lambda_{\sigma t}^{\star} ~t {} &
      {} & + {} &
      {} & \Lambda_{\sigma}^{\star} {} & {} & = {} & {} & 0 ~.
\end{alignat}
Here $\Lambda_{\diamond 3}$, $\Lambda_{\diamond 2}$, and $\Lambda_{\diamond 1}$
can be calculated using Eqs. (\ref{lambda3})-(\ref{lambda1}) with substituted
$C_{\text{c}}$ $=$ $I_{\text{c}}$ $=$ $O_{\text{c}}$ $=$ $U_{\text{c}}$ $=$ 0
and
\begin{align}\label{lambdavts}
\Lambda_{a t}^{\star} = - \left | \begin{array}{ccc}
      E_{\text{c}} & B_{\text{c}} & D_{\text{c}} \\
      K_{\text{c}} & H_{\text{c}} & J_{\text{c}} \\
      W_{\text{c}} & T_{\text{c}} & V_{\text{c}} \\
      \end{array} \right | ~, ~~
\Lambda_{e t}^{\star} = - \left | \begin{array}{ccc}
      A_{\text{c}} & E_{\text{c}} & D_{\text{c}} \\
      G_{\text{c}} & K_{\text{c}} & J_{\text{c}} \\
      S_{\text{c}} & W_{\text{c}} & V_{\text{c}} \\
      \end{array} \right | ~, ~~
\Lambda_{\sigma t}^{\star} = - \left | \begin{array}{ccc}
      A_{\text{c}} & B_{\text{c}} & E_{\text{c}} \\
      G_{\text{c}} & H_{\text{c}} & K_{\text{c}} \\
      S_{\text{c}} & T_{\text{c}} & W_{\text{c}} \\
      \end{array} \right | ~,
\end{align}
\begin{alignat}{8}\label{lambdavs}
\Lambda_{a}^{\star} {} & = {} & {} & - {} & {} & \left | \begin{array}{ccc}
      F_{\text{c}} & B_{\text{c}} & D_{\text{c}} \\
      L_{\text{c}} & H_{\text{c}} & J_{\text{c}} \\
      X_{\text{c}} & T_{\text{c}} & V_{\text{c}} \\
      \end{array} \right | {} & {} & - {} & {} & \left | \begin{array}{cc}
      B_{\text{c}} & E_{\text{c}} \\
      H_{\text{c}} & K_{\text{c}} \\
      \end{array} \right | {} & {} & - {} & {} & \left | \begin{array}{cc}
      D_{\text{c}} & E_{\text{c}} \\
      V_{\text{c}} & W_{\text{c}} \\
      \end{array} \right | ~, {} & {} &
\notag \\
\Lambda_{e}^{\star} {} & = {} & {} & - {} & {} & \left | \begin{array}{ccc}
      A_{\text{c}} & F_{\text{c}} & D_{\text{c}} \\
      G_{\text{c}} & L_{\text{c}} & J_{\text{c}} \\
      S_{\text{c}} & X_{\text{c}} & V_{\text{c}} \\
      \end{array} \right | {} & {} & - {} & {} & \left | \begin{array}{cc}
      E_{\text{c}} & A_{\text{c}} \\
      K_{\text{c}} & G_{\text{c}} \\
      \end{array} \right | {} & {} & - {} & {} & \left | \begin{array}{cc}
      J_{\text{c}} & K_{\text{c}} \\
      V_{\text{c}} & W_{\text{c}} \\
      \end{array} \right | ~, {} & {} &
\notag \\
\Lambda_{\sigma}^{\star} {} & = {} & {} & - {} & {} & \left | \begin{array}{ccc}
      A_{\text{c}} & B_{\text{c}} & F_{\text{c}} \\
      G_{\text{c}} & H_{\text{c}} & L_{\text{c}} \\
      S_{\text{c}} & T_{\text{c}} & X_{\text{c}} \\
      \end{array} \right | {} & {} & - {} & {} & \left | \begin{array}{cc}
      K_{\text{c}} & H_{\text{c}} \\
      W_{\text{c}} & T_{\text{c}} \\
      \end{array} \right | {} & {} & - {} & {} & \left | \begin{array}{cc}
      E_{\text{c}} & A_{\text{c}} \\
      W_{\text{c}} & S_{\text{c}} \\
      \end{array} \right | ~. {} & {} &
\end{alignat}
The general solution of Eqs. (\ref{notation}) is
\begin{equation}\label{cubicsolutions}
\delta_{\diamond} =
A_{\diamond 1}^{\star} \text{e}^{\lambda_{1} t} +
A_{\diamond 2}^{\star} \text{e}^{\lambda_{2} t} +
A_{\diamond 3}^{\star} \text{e}^{\lambda_{3} t} -
\frac{\Lambda_{\diamond t}^{\star}}{\Lambda_{1}} t +
\frac{\Lambda_{2} \Lambda_{\diamond t}^{\star} - \Lambda_{1}
\Lambda_{\diamond}^{\star}}{\Lambda_{1}^{2}} ~.
\end{equation}
Here $\lambda_{i}$ with $i$ $=$ 1, 2, 3 are roots of Eq. (\ref{cubic}) and
\begin{align}\label{Ais}
A_{\diamond 1}^{\star} &= \frac{\ddot{\delta}_{\diamond} (0) -
      \left ( \dot{\delta}_{\diamond} (0) +
      \frac{\Lambda_{\diamond t}^{\star}}{\Lambda_{1}} \right )
      \left ( \lambda_{2} + \lambda_{3} \right ) -
      \frac{\Lambda_{2} \Lambda_{\diamond t}^{\star} -
      \Lambda_{1} \Lambda_{\diamond}^{\star}}{\Lambda_{1}^{2}}
      \lambda_{2} \lambda_{3}}{\left ( \lambda_{1} - \lambda_{2} \right )
      \left ( \lambda_{1} - \lambda_{3} \right )} ~,
\notag \\
A_{\diamond 2}^{\star} &= \frac{\ddot{\delta}_{\diamond} (0) -
      \left ( \dot{\delta}_{\diamond} (0) +
      \frac{\Lambda_{\diamond t}^{\star}}{\Lambda_{1}} \right )
      \left ( \lambda_{1} + \lambda_{3} \right ) -
      \frac{\Lambda_{2} \Lambda_{\diamond t}^{\star} -
      \Lambda_{1} \Lambda_{\diamond}^{\star}}{\Lambda_{1}^{2}}
      \lambda_{1} \lambda_{3}}{\left ( \lambda_{1} - \lambda_{2} \right )
      \left ( \lambda_{3} - \lambda_{2} \right )} ~,
\notag \\
A_{\diamond 3}^{\star} &= \frac{\ddot{\delta}_{\diamond} (0) -
      \left ( \dot{\delta}_{\diamond} (0) +
      \frac{\Lambda_{\diamond t}^{\star}}{\Lambda_{1}} \right )
      \left ( \lambda_{1} + \lambda_{2} \right ) -
      \frac{\Lambda_{2} \Lambda_{\diamond t}^{\star} -
      \Lambda_{1} \Lambda_{\diamond}^{\star}}{\Lambda_{1}^{2}}
      \lambda_{1} \lambda_{2}}{\left ( \lambda_{1} - \lambda_{3} \right )
      \left ( \lambda_{2} - \lambda_{3} \right )} ~.
\end{align}
The constants in Eq. (\ref{cubicsolution}) and Eq. (\ref{cubicsolutions})
are related in such a way that
\begin{align}\label{relation}
A_{\diamond 1} = A_{\diamond 1}^{\star} \lambda_{1} ~, ~~
A_{\diamond 2} = A_{\diamond 2}^{\star} \lambda_{2} ~, ~~
A_{\diamond 3} = A_{\diamond 3}^{\star} \lambda_{3} ~, ~~
B_{\diamond} = \frac{\Lambda_{2} \Lambda_{\diamond t}^{\star} - \Lambda_{1}
\Lambda_{\diamond}^{\star}}{\Lambda_{1}^{2}} ~.
\end{align}
The evolution of longitude of pericenter is not ignored in the solution
given by Eq. (\ref{cubicsolution}) that includes $\delta_{\tilde{\omega}}$.

\section{Linearization parameters}
\label{app:lp}

This appendix presents parameters giving linearization solutions with
the best frequency accordance in Figs. \ref{fig:spap} and \ref{fig:mnap}.
In order to obtain the accuracy of parameters 5 valid places in the used
model we divided the synodic period during the averaging into a larger
number of equal steps as during the calculation of the linearization solutions
in Figs. \ref{fig:spap} and \ref{fig:mnap}. From the constant coefficients
in Eqs. (\ref{alphabet}) the coefficients that have the partial derivatives
of $R$ with respect to $\sigma$ are most sensitive on the number of steps,
particularly at the resonances with the Earth.

\begin{landscape}
\begin{table}
\caption{The parameters of linearization solution obtained for
the evolution 5 in Fig. \ref{fig:spap}. The synodic period is divided
into 10$^{8}$ equal steps during the averaging in order to obtain 5 valid
places. It is 10$^{4}$ times more steps as in Fig. \ref{fig:spap}.
Decimal digits shown in the parenthesis may have not been yet accurately
determined in the used model. 10$^{3}$ equal steps in the synodic period
is usually enough in order to obtain an usable linearization solution.}
\label{tab:splp}
\begin{center}
\begin{tabular}{| M M |}
\hline
a_{0} &= 1.1182 ~\text{AU} &
e_{0} &= 0.39994\\
\omega_{0} &= 0.48186 ~\text{rad} &
\sigma_{0} &= 2.4170 ~\text{rad}\\
A_{\text{c}} &= 3.5583 \times 10^{-5} ~\text{year}^{-1} &
B_{\text{c}} &= - 0.00030335 ~\text{AU}.\text{year}^{-1}\\
C_{\text{c}} &= 0 ~\text{AU}.\text{year}^{-1}.\text{rad}^{-1} &
D_{\text{c}} &= 0.00012517 ~\text{AU}.\text{year}^{-1}.\text{rad}^{-1}\\
E_{\text{c}} &= 0 ~\text{AU}.\text{year}^{-2} &
F_{\text{c}} &= - 5.338(6) \times 10^{-8} ~\text{AU}.\text{year}^{-1}\\
G_{\text{c}} &= 3.0867 \times 10^{-5} ~\text{AU}^{-1}.\text{year}^{-1} &
H_{\text{c}} &= - 0.00012580 ~\text{year}^{-1}\\
I_{\text{c}} &= 0 ~\text{year}^{-1}.\text{rad}^{-1} &
J_{\text{c}} &= 1.0673 \times 10^{-5} ~\text{year}^{-1}.\text{rad}^{-1}\\
K_{\text{c}} &= 0 ~\text{year}^{-2} &
L_{\text{c}} &= - 1.5564 \times 10^{-5} ~\text{year}^{-1}\\
M_{\text{c}} &= 0.00015918 ~\text{AU}^{-1}.\text{year}^{-1}.\text{rad} &
N_{\text{c}} &= 0.0031374 ~\text{year}^{-1}.\text{rad}\\
O_{\text{c}} &= 0 ~\text{year}^{-1} &
P_{\text{c}} &= - 2.2552 \times 10^{-5} ~\text{year}^{-1}\\
Q_{\text{c}} &= 0 ~\text{year}^{-2}.\text{rad} &
R_{\text{c}} &= - 4.7476 \times 10^{-5} ~\text{year}^{-1}.\text{rad}\\
S_{\text{c}} &= - 42.147 ~\text{AU}^{-1}.\text{year}^{-1}.\text{rad} &
T_{\text{c}} &= - 0.0024984 ~\text{year}^{-1}.\text{rad}\\
U_{\text{c}} &= 0 ~\text{year}^{-1} &
V_{\text{c}} &= 7.1559 \times 10^{-5} ~\text{year}^{-1}\\
W_{\text{c}} &= 0 ~\text{year}^{-2}.\text{rad} &
X_{\text{c}} &= 0.0035367 ~\text{year}^{-1}.\text{rad}\\
\Lambda_{3} &= 1.8651 \times 10^{-5} ~\text{year}^{-1} &
\Lambda_{2} &= 0.0052758 ~\text{year}^{-2}\\
\Lambda_{1} &= 5.2720 \times 10^{-7} ~\text{year}^{-3} &
\Lambda_{0} &= 0 ~\text{year}^{-4}\\
\Lambda_{a t} &= 0 ~\text{AU}.\text{year}^{-5} &
\Lambda_{e t} &= 0 ~\text{year}^{-5}\\
\Lambda_{\omega t} &= 0 ~\text{year}^{-5}.\text{rad} &
\Lambda_{\sigma t} &= 0 ~\text{year}^{-5}.\text{rad}\\
\Lambda_{a} &= 0 ~\text{AU}.\text{year}^{-4} &
\Lambda_{e} &= 0 ~\text{year}^{-4}\\
\Lambda_{\omega} &= 2.7804 \times 10^{-10} ~\text{year}^{-4}.\text{rad} &
\Lambda_{\sigma} &= 0 ~\text{year}^{-4}.\text{rad}\\
\lambda_{1} &= \left ( 4.0639 \times 10^{-5} +
0.072635 ~i \right ) ~\text{year}^{-1} &
\lambda_{2} &= - 9.9929 \times 10^{-5} ~\text{year}^{-1}\\
\lambda_{3} &= \left ( 4.0639 \times 10^{-5} -
0.072635 ~i \right ) ~\text{year}^{-1} &
 & \\
A_{a 1} &= \left ( - 2.707(7) \times 10^{-8} -
3.0800 \times 10^{-6} ~i \right ) ~\text{AU}.\text{year}^{-1} &
A_{a 2} &= 7.6882 \times 10^{-10} ~\text{AU}.\text{year}^{-1}\\
A_{a 3} &= \left ( - 2.707(7) \times 10^{-8} +
3.0800 \times 10^{-6} ~i \right ) ~\text{AU}.\text{year}^{-1} &
B_{a} &= 9.2501 \times 10^{-5} ~\text{AU}\\
A_{e 1} &= \left ( - 3.1277 \times 10^{-9} -
2.6261 \times 10^{-7} ~i \right ) ~\text{year}^{-1} &
A_{e 2} &= - 1.5558 \times 10^{-5} ~\text{year}^{-1}\\
A_{e 3} &= \left ( - 3.1277 \times 10^{-9} +
2.6261 \times 10^{-7} ~i \right ) ~\text{year}^{-1} &
B_{e} &= - 0.15568\\
A_{\omega 1} &= \left ( - 1.3289 \times 10^{-8} +
5.5509 \times 10^{-7} ~i \right ) ~\text{year}^{-1}.\text{rad} &
A_{\omega 2} &= 0.00047994 ~\text{year}^{-1}.\text{rad}\\
A_{\omega 3} &= \left ( - 1.3289 \times 10^{-8} -
5.5509 \times 10^{-7} ~i \right ) ~\text{year}^{-1}.\text{rad} &
B_{\omega} &= 4.8028 ~\text{rad}\\
A_{\sigma 1} &= \left ( 0.0017872 -
1.6473 \times 10^{-5} ~i \right ) ~\text{year}^{-1}.\text{rad} &
A_{\sigma 2} &= - 3.7704 \times 10^{-5} ~\text{year}^{-1}.\text{rad}\\
A_{\sigma 3} &= \left ( 0.0017872 +
1.6473 \times 10^{-5} ~i \right ) ~\text{year}^{-1}.\text{rad} &
B_{\sigma} &= - 0.37688 ~\text{rad}\\
\hline
\end{tabular}
\end{center}
\end{table}
\end{landscape}

\begin{landscape}
\begin{table}
\caption{Similarly as Table 2 but for the evolution 5 in
Fig. \ref{fig:mnap} in the problem without rotational symmetry.
The synodic period is divided into 10$^{6}$ equal steps during
the averaging in order to obtain 5 valid places. It is 10$^{2}$ times
more steps as in Fig. \ref{fig:mnap}. Decimal digits shown in the parenthesis
may have not been yet accurately determined in the used model.}
\label{tab:mnlp}
\begin{center}
\begin{tabular}{| M M |}
\hline
a_{0} &= 35.186 ~\text{AU} &
e_{0} &= 0.39994\\
\omega_{0} &= 1.9380 ~\text{rad} &
\sigma_{0} &= 3.8701 ~\text{rad}\\
A_{\text{c}} &= 2.4457 \times 10^{-7} ~\text{year}^{-1} &
B_{\text{c}} &= 1.9715 \times 10^{-5} ~\text{AU}.\text{year}^{-1}\\
C_{\text{c}} &= - 9.6498 \times 10^{-8} ~\text{AU}.\text{year}^{-1}.
\text{rad}^{-1} &
D_{\text{c}} &= 5.6697 \times 10^{-5} ~\text{AU}.\text{year}^{-1}.
\text{rad}^{-1}\\
E_{\text{c}} &= 0 ~\text{AU}.\text{year}^{-2} &
F_{\text{c}} &= - 2.2947 \times 10^{-7} ~\text{AU}.\text{year}^{-1}\\
G_{\text{c}} &= 1.2651 \times 10^{-9} ~\text{AU}^{-1}.\text{year}^{-1} &
H_{\text{c}} &= - 1.2283 \times 10^{-6} ~\text{year}^{-1}\\
I_{\text{c}} &= 9.0683 \times 10^{-7} ~\text{year}^{-1}.\text{rad}^{-1} &
J_{\text{c}} &= 4.6136 \times 10^{-7} ~\text{year}^{-1}.\text{rad}^{-1}\\
K_{\text{c}} &= 0 ~\text{year}^{-2} &
L_{\text{c}} &= - 1.4326 \times 10^{-7} ~\text{year}^{-1}\\
M_{\text{c}} &= 1.6801 \times 10^{-7} ~\text{AU}^{-1}.\text{year}^{-1}.
\text{rad} &
N_{\text{c}} &= 9.6325 \times 10^{-6} ~\text{year}^{-1}.\text{rad}\\
O_{\text{c}} &= 4.9475 \times 10^{-7} ~\text{year}^{-1} &
P_{\text{c}} &= 8.1150 \times 10^{-7} ~\text{year}^{-1}\\
Q_{\text{c}} &= 0 ~\text{year}^{-2}.\text{rad} &
R_{\text{c}} &= 1.4733 \times 10^{-6} ~\text{year}^{-1}.\text{rad}\\
S_{\text{c}} &= - 0.0032489 ~\text{AU}^{-1}.\text{year}^{-1}.\text{rad} &
T_{\text{c}} &= - 2.2703 \times 10^{-5} ~\text{year}^{-1}.\text{rad}\\
U_{\text{c}} &= - 8.6844 \times 10^{-7} ~\text{year}^{-1} &
V_{\text{c}} &= - 3.7309 \times 10^{-8} ~\text{year}^{-1}\\
W_{\text{c}} &= 0 ~\text{year}^{-2}.\text{rad} &
X_{\text{c}} &= - 9.0682 \times 10^{-6} ~\text{year}^{-1}.\text{rad}\\
\Lambda_{3} &= 5.2628 \times 10^{-7} ~\text{year}^{-1} &
\Lambda_{2} &= 1.8420 \times 10^{-7} ~\text{year}^{-2}\\
\Lambda_{1} &= 1.6444 \times 10^{-13} ~\text{year}^{-3} &
\Lambda_{0} &= - 1.6899 \times 10^{-18} ~\text{year}^{-4}\\
\Lambda_{a t} &= 0 ~\text{AU}.\text{year}^{-5} &
\Lambda_{e t} &= 0 ~\text{year}^{-5}\\
\Lambda_{\omega t} &= 0 ~\text{year}^{-5}.\text{rad} &
\Lambda_{\sigma t} &= 0 ~\text{year}^{-5}.\text{rad}\\
\Lambda_{a} &= - 2.8673 \times 10^{-21} ~\text{AU}.\text{year}^{-4} &
\Lambda_{e} &= - 2.5971 \times 10^{-19} ~\text{year}^{-4}\\
\Lambda_{\omega} &= - 1.3411 \times 10^{-19} ~\text{year}^{-4}.\text{rad} &
\Lambda_{\sigma} &= 9.6932 \times 10^{-20} ~\text{year}^{-4}.\text{rad}\\
\lambda_{1} &= 2.6152 \times 10^{-6} ~\text{year}^{-1} &
\lambda_{2} &= - 3.5079 \times 10^{-6} ~\text{year}^{-1}\\
\lambda_{3} &= \left ( 1.8318 \times 10^{-7} +
0.00042920 ~i \right ) ~\text{year}^{-1} &
\lambda_{4} &= \left ( 1.8318 \times 10^{-7} -
0.00042920 ~i \right ) ~\text{year}^{-1}\\
C_{a 1} &= - 0.00051231 ~\text{AU} &
C_{a 2} &= - 0.0005978(8) ~\text{AU}\\
C_{a 3} &= \left ( 0.0014034 + 0.00026880 ~i \right ) ~\text{AU} &
C_{a 4} &= \left ( 0.0014034 - 0.00026880 ~i \right ) ~\text{AU}\\
C_{e 1} &= 0.064938 &
C_{e 2} &= 0.088719\\
C_{e 3} &= 1.1419 \times 10^{-5} + 2.1838 \times 10^{-6} ~i &
C_{e 4} &= 1.1419 \times 10^{-5} - 2.1838 \times 10^{-6} ~i\\
C_{\omega 1} &= 0.28648 ~\text{rad} &
C_{\omega 2} &= - 0.20716 ~\text{rad}\\
C_{\omega 3} &= \left ( 2.0242 \times 10^{-5} +
3.0373 \times 10^{-6} ~i \right ) ~\text{rad} &
C_{\omega 4} &= \left ( 2.0242 \times 10^{-5} -
3.0373 \times 10^{-6} ~i \right ) ~\text{rad}\\
C_{\sigma 1} &= - 0.022115 ~\text{rad} &
C_{\sigma 2} &= - 0.031163 ~\text{rad}\\
C_{\sigma 3} &= \left ( - 0.0020403 + 0.010623 ~i \right ) ~\text{rad} &
C_{\sigma 4} &= \left ( - 0.0020403 - 0.010623 ~i \right ) ~\text{rad}\\
\hline
\end{tabular}
\end{center}
\end{table}
\end{landscape}

\begin{acknowledgements}
A part of this work was done at the Tekov Observatory but largest
pieces of understanding were found at different places. I would like to thank
the referees of this paper for their useful comments.
\end{acknowledgements}

Conflict of Interest: The author declares that he has no conflict
of interest.

\end{document}